\documentclass[prd,reprint,superscriptaddress,preprintnumbers,nofootinbib,amsmath,amssymb,aps]{revtex4-2}
\usepackage{graphicx}
\usepackage{multirow}
\usepackage{dcolumn}
\usepackage{color}
\usepackage[colorlinks=true,citecolor=blue,urlcolor=blue]{hyperref}
\usepackage{import}
\begin{document}
\title{Growth of matter perturbations in the Interacting Dark Energy/Dark Matter Scenarios\\}
\author{N.Nazari Pooya}
\thanks{Email: nazarip@basu.ac.ir}
\affiliation{Department of Physics, Faculty of Science, Bu-Ali Sina University, Hamedan
Iran}
\begin{abstract}
In this study, we investigate two widely recognized Interacting Dark Energy(IDE) models and assess their compatibility with observational data, focusing on the growth rate of matter perturbations. We explore IDE models with different equations of state (EoS) parameters for Dark Energy (DE), including the CPL parameterization and a constant value for $ w_{\mathrm{de}}$. To constrain the parameters of the IDE models using background data, we employ a Markov Chain Monte Carlo (MCMC) analysis. Our results show that both IDE-I and IDE-II models are Compatible with observational data, although with slight variations influenced by the homogeneity or clustering of DE. Following that, we investigate the growth of matter perturbations and perform a comprehensive statistical analysis utilizing both the background and growth rate data. The growth rate in IDE models exhibits deviations compared to the $\Lambda\mathrm{CDM}$ model due to the impact of homogeneity or clustering of DE, as well as the selection of the EoS parameter. However, we find that the IDE models show good compatibility with the growth rate data. Furthermore, we explore how the clustering or homogeneity of DE and the selection of the EoS parameter affect the evolution of the relative difference in the growth rate of IDE models, $\Delta f $, in comparison to the $\Lambda\mathrm{CDM}$ model.
Lastly, we employ the AIC and BIC criteria to evaluate and identify the best model that is compatible with the observational data. The selection of the model depends on the homogeneity or clustering of DE, the EoS parameter, and the dataset used. Overall, the IDE-I and IDE-II models exhibit agreement with the data, with slight deviations depending on specific scenarios and parameters.
\end{abstract}
%%%%%%%%%%%%%
\maketitle
%%%%%%%%%%%%%%%%%%%
\section{Introduction}
Observational evidence from various sources strongly supports the notion that the expansion of the Universe is accelerating. This evidence encompasses diverse measurements, including Supernovae type Ia (SnIa) \citep{Riess1998,Perlmutter1999,Kowalski2008}, Cosmic Microwave Background (CMB) \citep{Jarosik2011, Komatsu2011, Planck Collaboration XIV2016, Aghanim et al2020}, Baryon Acoustic Oscillations (BAO) \citep{Reid2012,Blake2011b,Percival2010,Tegmark2004,Cole2005}, high redshift galaxy clusters \citep{WangSteinhardt1998a,Allen2004}, weak lensing surveys \citep{Fu:2008,Amendola:2007rr,Benjamin:2007}, and other sources. These diverse observations consistently support the idea of accelerated expansion, shedding light on the evolution of the Universe and the role of Dark Energy (DE).
However, despite the strong support for the $\Lambda \mathrm{CDM}$ model provided by these observations, several challenges persist. The nature of DE itself remains a mystery, as its origin and properties are not yet fully understood. The cosmological constant's fine-tuning and the cosmic coincidence problem raise questions about why DE dominates the Universe's energy density at the present epoch. Moreover, tensions related to $S_{8}$ \citep{Joudaki2018,Abbott2018,Basilakos2017} and $\mathrm{H}_{0}$ \citep{Abdalla2022,Kazantzidis2019,Valentino2021,Schoneberg2022,Alestas2020,Shah2021} further complicate matters, posing both theoretical and observational challenges.

Observationally, there are discrepancies in the formation of cosmic structures at smaller scales compared to the predictions of the $\Lambda \mathrm{CDM}$ model. These inconsistencies and tensions necessitate exploring alternative theories and modifications to address these issues and refine our understanding expansion of the Universe. Scientists are actively investigating various approaches, including modified gravity theories \citep{Capozziello2002, Carroll2004,Nicolis2009, Deffayet2009,Dvali2000,Nojiri2005,Koivisto2007}, Unified DE models \citep{Farnes2018,Cardone2004,Paliathanasis2023,Ansoldi2013,Bousder2020}, and Interacting DE (IDE) models \citep{Liu2022,Guo2005,Cai2002,Bi2005,Ferreira2017,Landim2019,Vagnozzi2020,Costa2014},
and many other alternative cosmological scenarios,
to overcome these theoretical and observational challenges.

IDE models, which involve the interaction between DE and Dark Matter(DM), have significant implications for the evolution of the Universe and the behavior of these enigmatic components. Extensive research has been conducted on these models to comprehend their impact on the expansion history of the Universe, the growth of large-scale structures, and observational constraints. Since the exact form of the interaction cannot be deduced from fundamental principles due to the unknown nature of DE and DM, a phenomenological approach is often employed to determine the nature of their interaction. Recent observational data indicates that a direct interaction between DE and DM cannot be ruled out. These models introduce additional parameters to describe the strength of the interaction and its effects on observables related to large-scale structures.

The growth of matter perturbations in the Universe can be influenced by DE through various mechanisms, even without considering the interaction between two dark sectors. One such mechanism is the deceleration of the growth rate due to the accelerating expansion of the Universe, resulting in a slower evolution of matter perturbations. Additionally, DE can exhibit perturbations that grow in a similar manner to DM, leading to changes in the distribution and clustering of DM throughout the Universe. These mechanisms demonstrate how DE can impact the growth of matter perturbations in the Universe\citep{Pace2010,Garriga1999,Abramo2008,Tegmark2004,Sapone2012,Basse2014,Nesseris2015,Mota2007, Dossett2013,Batista2013,Batista2014,Ballesteros2008}.

The growth of DM perturbations in IDE models can be influenced by perturbations in DE, where an exchange of energy between the two dark components impacts their evolutions. To comprehend the growth of structures caused by this interaction, researchers utilize theoretical modeling, simulations, and analysis of observational data. These investigations significantly advance our understanding of the Universe's evolution by uncovering the interaction between DE  and DM. 

This study aims to examine the effects of small DE perturbations on the growth of DM perturbations in IDE models. By studying the growth of DM perturbations in the presence of DE perturbations, valuable insights are gained into the complex interaction between these dark components and their influence on the evolution of cosmic structures.
To explore the impact of DE perturbations on the growth of DM perturbations, we consider different interaction terms and Equation of State (EoS) parameters for DE in the conservation equations related to DE and DM. Subsequently, we solve the coupled equations governing the evolution of DE and DM, and finally, we compare the resulting outcomes with observational data. This analysis enables us to measure and evaluate any deviations between the predictions of the standard $ \Lambda$CDM model and the observational data.
 
 The article's structure is outlined as follows:
In Sec. \ref{sec2}, we derive the necessary equations for the background evolution of the Universe and introduce the IDE models investigated in this study.
In Sec. \ref{sec3}, we give a brief overview of the current observational datasets at the background level. We then utilize numerical Markov Chain Monte Carlo (MCMC) analysis to constrain the free parameters of the IDE models examined in this research.
In Sec. \ref{sec4}, we establish the fundamental equations governing the evolution of DE and DM in the linear regime within the IDE model scenarios. We also investigate the growth rate of matter perturbations in this section. Additionally, we incorporate growth rate data along with background data to obtain more comprehensive constraints on the free parameters of models.
In Sec. \ref{sec5}, we perform a comparison between the IDE models and the $ \Lambda $CDM model, using significant cosmological quantities. This comparison is performed using the best-fit values obtained from the likelihood analysis for the free parameters.
Finally, in Sec. \ref{conclude}, we present the conclusions derived from our study.
\section{Basic equations in IDE models: background level}\label{sec2}
In this study, the background of the Universe is described using the Friedmann-Robertson-Walker (FRW) metric. The FRW metric, which is defined in terms of conformal time, can be expressed as $ ds^{2}=a^{2}(\eta)\big (-\mathrm{d}\eta^{2}+\delta_{ij}\mathrm{d}\mathrm{x^{i}\mathrm{d} x^{j}} \big) $. 
where $ a(\eta) $ is the scale factor.
Also, the energy-momentum tensor of DE and DM, denoted by ${\bar{T}}_{\mu\nu} =\bar{p}\bar{g}_{\mu\nu}(\bar{p}+\bar{\rho})\bar{u}_{\mu}\bar{v}_{\nu}$. where the bars denote that the quantities are unperturbed. In the context of IDE models, the conservation equations for the energy-momentum of these components can be expressed as follows \citep{Marcondes2016}:
\begin{align}
&\nabla_{\mu}\bar{T}_{\;\; \nu}^{\mu ,{i}}=\bar{Q}^{{\;i}}_{\nu} 
\end{align}
where $ i= \mathrm{de}, \mathrm{dm}$, and $ \bar{Q}^{\;i}_{\nu} $ is the phenomenological interaction term among the DM and DE. Due to the conservation of total energy-momentum, we can conclude $ \bar{Q}^{\mathrm{\;de}}_{\nu}=-\bar{Q}^{\;\mathrm{dm}}_{\nu} $.
Also, in the case of non-interaction between DE and DM, we have $\bar{Q}^{\mathrm{\;de}}_{\nu} = -\bar{Q}^{\;\mathrm{dm}}_{\nu} = 0$. This means that the energy transfer rate from DE to DM or vice versa is zero.
Furthermore, because of the homogeneous and isotropic of background, the spatial components of $ \bar{Q}^{\;i}_{\nu}$ are zero. Consequently, the evolution of the energy density of dark energy (${\rho}_{{\mathrm{de}}}$), dark matter (${\rho}_{{\mathrm{dm}}}$), baryons (${\rho}_{{\mathrm{b}}}$), and radiation (${\rho}_{{\mathrm{r}}}$) over time can be determined using the following conservation equations.
\begin{align}
&\dot{{\rho}}_{\mathrm{dm}}+3\mathcal{H}{\rho}_{{\mathrm{dm}}}=\bar Q_{0}   \label{qdw} \\ 
&\dot{{\rho}}_{{\mathrm{de}}}+3\mathcal{H}(1+w_{\mathrm{de}}){\rho}_{{\mathrm{de}}}=-\bar Q_{0} \label{qsw}\\
&\dot{{\rho}}_{\mathrm{b}}+3\mathcal{H}{\rho}_{{\mathrm{b}}}=0\label{qhdw} \\ 
&\dot{{\rho}}_{\mathrm{r}}+4\mathcal{H}{\rho}_{{\mathrm{r}}}=0\label{qldw}  
\end{align}
where dots denote derivative with respect to the conformal time, $ \mathcal{H}$ is the conformal Hubble parameter($ \mathcal{H}=aH$), and $ w_{\mathrm{de}}=p_{\mathrm{de}}/\rho_{\mathrm{de}} $ is the EoS parameter of DE.
Moreover, the evolution of a spatially flat FRW Universe with a homogeneous and isotropic background is governed by the following equation:
\begin{equation}\label{etqq} 
\mathcal{H}^{2}=\frac{8\pi \mathrm{G}}{3}a^{2}(\rho_{\mathrm{dm}}+\rho_{\mathrm{de}}+\rho_{\mathrm{b}}+\rho_{\mathrm{r}}) 
\end{equation}
The solutions of Eqs. (\ref{qdw} \& \ref{qsw}) depend on the particular forms of $ \bar Q_{0}$ and $ w_{\mathrm{de}}$.
In this study, two phenomenological interaction terms for $\bar Q_{0}$ are considered:
$\bar Q_{01}=\xi_{1}\mathcal{H}\rho_{\mathrm{de}}$ and $\bar Q_{02}=\xi_{2}\mathcal{H}\rho_{\mathrm{dm}}$.
Where $\xi_{1}$ and $ \xi_{2}$ are dimensionless coupling parameter describing the strength of interaction between DE and DM. 
In recent years, there has been a significant amount of research devoted to models resembling these, in which the interaction term exhibits proportionality to the energy densities ($\rho_{\mathrm{de}}$, $\rho_{\mathrm{dm}}$), or a combination of both\citep{Abdalla2009,Abdalla2010,Cao2011,BWang2016}. 

Moreover, we investigate two distinct cases for the EoS parameter associated with each of the interaction terms. In the first case, we use a well-known parameterization called the Chevallier-Polarski-Linder (CPL) parameterization, defined as $w_{\mathrm{de}} = w_{0} + w_{1}(1-a)$\citep{PolarskiChevallier2001,Linder2003}. In the second case, we assume that the parameter $w_{\mathrm{de}}$ is constant. In the following, we examine these contents in more detail.
 \begin{table}
\centering
\caption[Equations of state]{ Phenomenological interaction models and their related equations of state that have been investigated in this research.}
 \begin{tabular}{lcccccc}
 \hline
 \hline
 Model  & $\bar{Q_{0}} $ &$ \bar{Q_{0}}/\rho_{\mathrm{de}} $ & $ \bar{Q_{0}}/\rho_{\mathrm{dm}} $&\;\; DE  EoS & \\
\hline
IDE I  \;\;\;\;\   &  $\mathcal{H}\xi_{1}\rho_{\mathrm{de}}$ & $\mathcal{H}\xi_{1}$ & \;\; $\mathcal{H}\xi_{1}\frac{\Omega_{\mathrm{de}}}{\Omega_{\mathrm{dm}}}$ &\;\;$\mathrm{CPL  \; , \; w_{\mathrm{d}}} $& \\
\hline
IDE II  \;\;\;\;\  & $\mathcal{H}\xi_{2}\rho_{\mathrm{dm}}$ & \;\;\;\; $\mathcal{H}\xi_{2}\frac{\Omega_{\mathrm{dm}}}{\Omega_{\mathrm{de}}}$& $\mathcal{H}\xi_{2}$&\;\;\;$ \mathrm{CPL \; , \; w_{\mathrm{d}}} $& \\
\hline
\hline\hline
\label{tab1222}
\end{tabular}
\end{table}

\textbf{Interaction Term I:} $\bar{Q}_{01}=\xi_{1}\mathcal{H}\rho_{\mathrm{de}}$:
By employing this interaction term and assuming the CPL  parameterization for DE in equations (\ref{qdw} \& \ref{qsw}), we can obtain the solutions for these equations as follows:
\begin{align}\notag
& \rho_{\mathrm{dm}}=\rho^{0}_{\mathrm{de}}\xi_{1} a^{-3}\int_{1}^{a}e^{3w_{1}(a-1)}a^{-3(w_{0}+w_{1})-\xi_{1} -1}\mathrm{d}\mathrm{a} \\
&\;\;\;\;\;\;\; +\rho^{0}_{\mathrm{dm}}a^{-3}\;\;\;   ;\\
&\rho_{\mathrm{de}}\; =\rho^{0}_{\mathrm{de}}a^{-3(1+w_{0}+w_{1})-\xi_{1}}e^{{3(a-1)w_{1}}}
 \end{align}
Furthermore, if we assume that $w_{\mathrm{de}}$ is a constant, we can determine the solutions for equations (\ref{qdw} \& \ref{qsw}) as below:
\begin{align}\notag
&\rho_{\mathrm{dm}}=\frac{\mathrm{a}^{-3((1+w_{\mathrm{de}})-\xi_{1}}}{3w_{\mathrm{de}}+\xi_{1}}\Big[\xi_{1}\rho_{\mathrm{de}}^{0}(\mathrm{a}^{3w_{\mathrm{de}} +\xi_{1}}-1)\\
& \;\;\;\; \;\;\; +(3w_{\mathrm{de}}+\xi_{1})\rho_{\mathrm{dm}}^{0}\mathrm{a}^{3w_{\mathrm{de}}+\xi_{1}}\Big]\\
&\rho_{\mathrm{de}}\;=\rho_{\mathrm{de}}^{0}\mathrm{a}^{-3((1+w_{\mathrm{de}})-\xi_{1}}
 \end{align}

\textbf{Interaction Term II:}\;\;$\bar{Q}_{02}=\xi_{2}\mathcal{H}\rho_{\mathrm{dm}}
 $:
By assuming this form of the interaction term and utilizing the CPL parameterization for DE in equations (\ref{qdw} \& \ref{qsw}), we can derive the solutions for these equations as follows:
\begin{align}
&\rho_{\mathrm{dm}}=\rho^{0}_{\mathrm{dm}}a^{-3+\xi_{2}}\;\; ;\\ \notag
&\rho_{\mathrm{de}}\;=a^{-3(1+w_{0}+w_{1})}\Big[\rho^{0}_{\mathrm{de}} e^{3(\mathrm{a}-1)w_{1}}-\rho^{0}_{\mathrm{dm}}e^{3aw_{1}} \\ 
&\;\;\;\;\;\; \times \xi_{2}  \int_{1}^{\mathrm{a}}e^{-3aw_{1}}a^{3(w_{0}+w_{1})+\xi_{2}-1}\mathrm{d}\mathrm{a}\Big]
 \end{align}
when $w_{\mathrm{de}}$ is constant, the solutions to Eqs.(\ref{qdw} \& \ref{qsw}) can be derived as follows.
 \begin{align}\notag
 &\rho_{\mathrm{dm}}\;=\rho_{\mathrm{dm}}^{0}\mathrm{a}^{-3+\xi_{2}}\\
&\rho_{\mathrm{de}}=\frac{\mathrm{a}^{-3(1+w_{\mathrm{de}})}}{3w_{\mathrm{de}}+\xi_{2}}\Big[ 
\rho_{\mathrm{de}}^{0}(3w_{\mathrm{de}}+\xi_{2})+\xi_{2}\rho_{\mathrm{dm}}^{0}(1-\mathrm{a}^{3w_{\mathrm{de}}+\xi_{2}}) \Big]
 \end{align}
In the following section, we analyze the observational constraints that are imposed on these IDE models at the background level.
\section{data analysis: background level}\label{sec3}
In this section, we provide an overview of the steps in analyzing observational data. 
The steps are as follows:

\textbf{I}. Background Analysis: The total likelihood at the background level is calculated using the $\chi^2_{\text{bac}}$ equation, which combines the contributions from different observational data sets, which is given by:
 \begin{equation}\label{xi22}
 \chi^2_{\text{bac}}(\mathbf{p}) = \chi^2_{\text{H}}(\mathbf{p}) + \chi^2_{\text{BAO}}(\mathbf{p}) + \chi^2_{\text{SN}}(\mathbf{p}) + \chi^2_{\text{CMB}}(\mathbf{p})
 \end{equation}
where, $\mathbf{p} = \{\Omega_{b}h^{2}, \Omega_{c}h^{2}, H_{0}, w_{0}, w_{1}, \xi_{1}, \xi_{2}\}$ represents the free parameters of the models, and the subscripts H, BAO, SN, and CMB denote the contributions from the Hubble parameter, Baryon Acoustic Oscillations, Supernova type Ia, and Cosmic Microwave Background, respectively. In this analysis, we utilize 1098 observational data points related to the background. These data points consist of 1048 data points from the Pantheon catalog for supernova type Ia (SnIa), 3 data points for the CMB, 11 data points for the BAO, and 36 data points for the H(z).

\textbf{II}. Growth Rate Analysis: In the second analysis, the growth rate data is incorporated. The total $\chi^2$ value ($\chi^2_{\text{tot}}$) combining the background and growth rate components is given by:
     \begin{equation}\label{xi222}
     \chi^2_{\text{tot}}(\mathbf{q}) = \chi^2_{\text{bac}}(\mathbf{p}) + \chi^2_{\text{growth}}
     \end{equation}
where, $\mathbf{q} = \{\mathbf{p}, \sigma_{8,0}\}$ represents the free parameters of the models at both the background and perturbation levels. The $\chi^2_{\text{growth}}$ term represents the contribution from the growth rate data. In this step, 44 growth rate data points are added to the background data points. Additionally, the details of $\chi^{2}_{\text{growth}}$ are explained in Subsec. \ref{growth}.

\textbf{III}. Statistical Tools: The $\chi^2$ statistic is commonly used to assess the level of agreement between theoretical models and observational data. Therefore, we utilize MCMC analysis, which explores the parameter space to determine uncertainties and correlations among the parameters. These statistical techniques are employed to analyze observational data and constrain model parameters based on their compatibility with the observational data. In the subsequent sections, we present a concise description of the datasets employed in this study.
   
\subsection{Type Ia Supernovae(SnIa) data}
The dataset of Type Ia Supernovae (SnIa) plays a crucial role in studying the dynamic background of the Universe and continues to provide valuable constraints for DE models. The SnIa dataset involves comparing the apparent magnitude with the absolute magnitude of observed SnIa, which is known as the distance modulus and theoretically is given by:
\begin{equation}
\mu_{th}(z)=5\log_{10}{d_{L}(z)}+ 42.384 - 5\log _{10}\mathrm{h}
%\mu_{0}
\end{equation}
Where, $d_{L}(z)$ represents the luminosity distance, which is defined in the following manner:
\begin{equation}
d_{L}(z)=\frac{c}{H_{0}}(1+z)\int_{0}^{z}\frac{dz^{\prime}}{E(z^{\prime})}
\end{equation}
In this analysis, we employ the Pantheon SnIa dataset, containing 1048 data points sourced from the Pantheon sample \citep{Scolnic2018}. Additionally, we obtain the respective $ \chi^{2}_{\mathrm{SN}} $ using the following relation:
\begin{equation}
\chi^{2}_{\mathrm{sn}}(\textbf{p})=\sum_{i=1}^{1048}\frac{[\mu_{th}(\textbf{p}, z_{i})-\mu_{obs}(z_{i})]^{2}}{\sigma_{\mu, i}^{2}}
\end{equation}
Where, the $\mu_{th}(\textbf{p}, z_{i})$ refers to the theoretical prediction of the distance modulus at a specific redshift $z_{i}$. On the other hand, $\mu_{obs}(z_{i})$ represents the distance modulus determined through observations and, $\sigma_{\mu, i}$ indicates the uncertainty related to the observational data.

\subsection{Baryon Acoustic Oscillations(BAO) data}
Recent investigations have highlighted the significance of BAO as a valuable geometric probe for examining DE. The precise position of the BAO peak in the CMB power spectrum is indeed dependent on the ratio of $ D_{V}(z) $ to the comoving sound horizon size $r_{s}(z) $ at the drag epoch, denoted as $ z_{d} $ which represents the epoch when baryons decoupled from photons.
In their study, Komatsu et al.\citep{Komatsu2008} noted that the drag epoch, characterized by $z_d$, occurs slightly later than the epoch of photon decoupling, represented by $z_*$. During this epoch, the gravitational potential well affects the behavior of baryons. As a result, the sound horizon size during the drag era is slightly larger compared to the photon decoupling era. Various researchers have reported their measurements of the BAO feature using different observable quantities.
Some measurements included constraints on the ratio $ r_{s}(z_{d})/D_{V}(z) $ or its inverse. The comoving sound horizon $r_{s}(z_{d}) $ is given by\citep{Komatsu2008}
\begin{equation}\label{rsa}
r_{s}(z_{d})=\frac{c}{H_{0}}\int^{\infty}_{z_{d}}\frac{c_{s}(z^{\prime})dz^{\prime}}{E(z^{\prime})}
\end{equation}
where $ c_{s}(z)={1}/[{{3}(1+\frac{3\Omega_{b0}}{4(1+z)\Omega_{\gamma 0}})}]^{\frac{1}{2}} $ and $ E(z) $ is given by Eq. (\ref{etqq}).
We adopt the approximate function for $z_d$ as described in \citep{EisensteinHu (1998)}. Furthermore, we set $\Omega_{\gamma0}=2.469\times10^{-5}h^{-2}$ (for $T_{\mathrm{cmb}}$ = 2.725 K) according to \citep{Hinshaw2013,Komatsu2008}. Also, the expression for $D_V(z)$ is provided in\citep{Komatsu2008} as follows:
\begin{equation}
D_{V}(z)=\frac{c}{H_{0}}\Big[(1+z)^{2}D_{A}^{2}(z)\frac{z}{E(z)}\Big]^{\frac{1}{3}}
\end{equation}
where $D_{A}(z)$ is the angular diameter distance. When the curvature density, $\Omega_{K}$, is zero, we can calculate $D_{A}(z)$ by using the following formula \citep{Komatsu2008}:
\begin{align}\label{daaa}
&D_{A}(z)=\frac{c}{H_{0}(1+z)}\int_{0}^{z}\frac{dz}{E(z)}
\end{align}
We utilize two datasets, one in the old format presented in Table \ref{tab1bao} and the other in the new format shown in Table \ref{tab2bao}. Since the data points listed in Tables \ref{tab1bao} and \ref{tab2bao} are uncorrelated, we calculate $\chi^{2}_{\mathrm{bao,1}}$ for the first case as follows:
\begin{equation}\label{bao1}
\chi^{2}_{\mathrm{bao,1}}=\sum_{i=1}^{4}\frac{[d_{z}(z_{i})|_{th}-(d_{z,i})|_{obs}]^{2}}{\sigma_{i}^{2}}
\end{equation}
In this case, the theoretical prediction is expressed as $d_{z}(z)=\frac{r_{s}(z_{d})}{D_{V}(z_{\mathrm{eff}})}$, where $r_{s}(z_{d})$ represents the comoving sound horizon size at the drag epoch, and $D_{V}(z_{\mathrm{eff}})$ denotes the effective volume distance. In the second case, $\chi^{2}_{\mathrm{bao,2}}$ is obtained from the following relation.
\begin{equation}\label{bao2}
\chi^{2}_{\mathrm{bao,2}}=\sum_{i=1}^{4}\frac{[\beta^{*}(z_{i})|_{th}-\beta^{*}_{z,i}|_{obs}]^{2}}{\sigma_{i}^{2}}
\end{equation}
In this case, the theoretical prediction is represented by $\beta^{*}(z)=\frac{D_{V}(z_{\mathrm{eff}})}{r_{s}(z_{d})}r_{s}^{\mathrm{fid}}$. Consequently, the total $\chi^{2}_{\mathrm{bao}}$ is given by $ \chi^{2}_{\mathrm{bao}}=\chi^{2}_{\mathrm{bao,1}}+\chi^{2}_{\mathrm{bao,2}} $.
\begin{table}
 \centering
 \caption[Equations of state]{The old format of the BAO data points, along with their Survey details and References.}
\begin{tabular}{lccccc} 
\hline
\hline
 $ z_{\mathrm{eff}} $  &$ d_{z}$  & Survey  &  Refs.\\
\hline\hline
 0.106& $0.336 \pm0.015$&$\mathrm{6dFGS} $ & \citep{Beutler2011}\\
 \hline
0.60&$ 0.0692\pm0.0033  $&$ \mathrm{WiggleZ} $& \citep{Blake2011} \\
 \hline
0.57&      $ 0.073 \pm.022 $   &   $\mathrm{SDSS.DR9} $  & \citep{Anderson2012} \\
 \hline
0.275&$  0.1390\pm0.0037 $&$\mathrm{SDSS.DR7} $&\citep{Percival2010} \\
\hline
 \label{tab1bao}
 \end{tabular}
 \begin{flushleft}
  \vspace{-0.5cm}
  {\small}
 \end{flushleft}
\end{table}
\begin{table}
 \centering
 \caption[Equations of state]{The new format of the BAO data points, along with their Survey details and References.}
\begin{tabular}{lccccc} 
\hline
\hline
  $  z_{\mathrm{eff}} $  &$\beta_{i}^{*}(\mathrm{Mpc})$  &$ r_{\mathrm{s}}^{\mathrm{fid}} $& Survey  &  Refs.\\
\hline\hline
 0.38& $1477 \pm16$&147.78& $\mathrm{BOSS.DR12} $ & \citep{Alam2017}\\
 \hline
0.51& $ 1877 \pm19$&147.78 &$\mathrm{BOSS.DR12} $ & \citep{Alam2017}\\
 \hline
0.61&$ 2140\pm22  $&147.78 &$ \mathrm{BOSS.DR12} $& \citep{Alam2017} \\
\hline
 0.15&$664\pm25  $&148.69 &$ \mathrm{BOSS.MGS} $& \citep{Ross2015} \\
 \hline
 \hline
 \label{tab2bao}
 \end{tabular}
 \begin{flushleft}
  \vspace{-0.5cm}
  {\small}
 \end{flushleft}
\end{table}
\subsection{\textbf{ Cosmic Microwave Background(CMB) data}}
The location of the CMB acoustic peak is valuable tool for constraining models of  DE as it depends on the angular diameter distance in dynamical DE models. The specific position of this peak in the power spectrum of temperature anisotropy in the CMB is determined by three parameters: $l_a$, $R$, and $\Omega_{b}h^{2}$.
Where, $l_{a}$ represents the angular scale of the sound horizon at the decoupling era, which can be calculated using the equation:
\begin{equation}
l_{a}=(1+z_{*})\frac{\pi D_{A}(z_{*})}{r_{s}(z_{*})}
\end{equation}
In this equation, $z_{*}$ refers to the redshift at the decoupling time, and a fitting formula from Hu\citep{Hu1996} is used to determine it. The coefficient of $(1+z_{*})$ is included because $D_{A}(z_{*})$ represents the physical angular diameter distance (see Eq. \ref{daaa}), while $r_{s}(z_{*})$ represents the comoving sound horizon at $z_{*}$ (see Eq. \ref{rsa}).
The scale distance or shift parameter at the decoupling epoch, denoted as $R$, is defined as  follows\citep{Bond1997}:
\begin{equation}
R(z_{*})=\frac{1}{c}\sqrt{\Omega_{m0}}H_{0}(1+z_{*})D_{A}(z_{*})
\end{equation}
Chen et al. \citep{Chen et al. 2019} conducted a comparison between the full CMB power spectrum analysis and the distance prior method to constrain different DE models. The results of both methods were found to be completely consistent. Therefore, in this study, we utilize the combined CMB likelihood (Planck 2018 TT, TE, EE + lowE) based on the observed values $X^{\mathrm{obs}}_{i} = \lbrace R, l_{a}, \Omega_{b}h^{2} \rbrace = \lbrace 1.7493, 301.462, 0.02239 \rbrace$, as obtained by Chen et al. \citep{Chen et al. 2019}. The $\chi^{2}_{\mathrm{cmb}}$ is expressed as follows:
\begin{equation}
\chi^{2}_{\mathrm{cmb}}=\Delta X_{i}\Sigma_{ij}^{-1}\Delta X_{i}^{T}
\end{equation}
Where, $\Delta X_{i}=\lbrace X^{\mathrm{th}}_{i}-X_{i}^{\mathrm{obs}}\rbrace$ represents the difference between the theoretical value $X^{\mathrm{th}}_{i}$ and the observed value $X_{i}^{\mathrm{obs}}$. The inverse of the covariance matrix $\Sigma_{ij}^{-1}$ associated with $\Delta X_{i}$ is given by:
\begin{equation*}
\Sigma_{ij}^{-1} =\left(
\begin{matrix}
94392.3971 & -1360.4913 &1664517.2916 \\
-1360.4913 & 161.4349 & 3671.6180 \\
1664517.2916&3671.6180& 79719182.5162
\end{matrix}
\right)
\end{equation*}
\subsection{\textbf{Hubble data}}
In our analysis, we utilize 36 data points of $H(z)$ from Table (\ref{tabHdata}), spanning the redshift range $0.07 \leqslant z \leqslant 2.34$. Since the measurements of $H(z)$ are uncorrelated, we can express the $\chi^{2}_{H}$ statistic as follows:
\begin{equation}
\chi^{2}_{H}(\textbf{p}) = \sum_{i=1}^{36} \frac{[H_{th}(\textbf{p}, z_{i}) - H_{obs}(z_{i})]^{2}}{\sigma^{2}_{i}}
\end{equation}
Here, $H_{th}(\textbf{p}, z_{i})$ represents the model predictions at the redshift $z_{i}$, while $H_{obs}(z_{i})$ and $\sigma_{i}$ denote the measured values and Gaussian errors, respectively, corresponding to the data points listed in Table (\ref{tabHdata}).
\begin{table}
\centering
\caption{The H(z) data used in the current analysis (in units of  $ \mathrm{km\; s^{-1}Mpc^{-1}} $). This compilation is partly based on those of Ref  \citep{Moresco:2016mzx}.}
\begin{tabular}{lclc|lclcccc} 
		\hline \hline
$ z $ & H(z)\;\;&$\sigma_{H}$\;\;\; & Refs. & $ z$ & H(z)\;\;&$ \sigma_{H} $\;\;\; & Refs.  \\
		\hline
		  $0.07$ & $69 $ & 19.6& \citep{Zhang:2012mp} &$0.48$ & $97 $&62 & \citep{Stern:2009ep}  \\

		 $0.09$ & $69$&12 & \citep{Stern:2009ep} & $0.57$ & $96.8 $ & 3.4& \citep{Anderson:2013zyy}  \\

		  $0.12$ & $68.6 $& 26.2 & \citep{Zhang:2012mp} & $0.593$ & $104$& 13 & \citep{Moresco_2012}  \\

		 $0.17$ & $83$ &  8& \citep{Stern:2009ep} & $0.6$ & $87.9$&  6.1 & \citep{Blake:2012pj}  \\
	
		 $0.179$ & $75$&  4 & \citep{Moresco_2012} &  $0.68$ & $92 $& 8 & \citep{Moresco_2012}  \\
	
		 $0.199$ & $75$&  5 & \citep{Moresco_2012} & $0.73$ & $97.3 $& 7 & \citep{Blake:2012pj}  \\
		
		  $0.2$ & $72.9$&  29.6 & \citep{Zhang:2012mp} & $0.781$ & $105 $& 12 & \citep{Moresco_2012}  \\

		  $0.27$ & $77 $& 14 & \citep{Stern:2009ep} & $0.875$ & $125 $& 17 & \citep{Moresco_2012}  \\

		  $0.28$ & $88.8$ & 36.6& \citep{Zhang:2012mp} & $0.88$ & $90$&  40 & \citep{Stern:2009ep}  \\
	
		  $0.35$ & $82.7 $& 8.4 & \citep{Chuang:2012qt} &  $0.9$ & $117$ & 23& \citep{Stern:2009ep}  \\
	
		  $0.352$ & $83 $& 14 & \citep{Moresco_2012} &  $1.037$ & $154$& 20 & \citep{Moresco_2012}  \\

		 $0.3802$ & $83$&  13.5 & \citep{Moresco:2016mzx} &  $1.3$ & $168$& 17 & \citep{Stern:2009ep}  \\
	
		  $0.4$ & $95$ & 17& \citep{Stern:2009ep} & $1.363$ & $160 $& 33.6 & \citep{Moresco:2015cya}  \\
	
		  $0.4004$ & $77$& 10.2 & \citep{Moresco:2016mzx} &  $1.43$ & $177 $& 18 & \citep{Stern:2009ep}  \\
	
		  $0.4247$ & $87.1$ &  11.2& \citep{Moresco:2016mzx} & $1.53$ & $140$& 14 & \citep{Stern:2009ep}  \\

		  $0.44$ & $82.6$&  7.8 & \citep{Blake:2012pj} &  $1.75$ & $202 $& 40 & \citep{Stern:2009ep}  \\
	
		 $0.44497$ & $92.8$& 12.9 & \citep{Moresco:2016mzx} &$1.965$ & $186.5 $& 50.4 & \citep{Moresco:2015cya}  \\

	    $0.4783$ & $80.9 $& 9 & \citep{Moresco:2016mzx} &  $2.34$ & $222$&7 &  \citep{Font-Ribera:2013wce}  \\
			\hline \hline
	\end{tabular}\label{tabHdata}
\end{table}
\section{Basic equations in IDE models: perturbations level}\label{sec4}
The perturbed Friedmann-Robertson-Walker (FRW) metric is used to describe the spacetime geometry in cosmology, taking into account small perturbations from the homogeneous and isotropic background Universe. In the conformal Newtonian gauge, the metric can be expressed as follows:
\begin{equation}
\mathrm{d}\mathrm{s}^2 =\mathrm{a}(\eta)^{2}[ - (1+2\psi)d\eta^{2} + (1-2\phi)\delta_{\mathrm{ij}} \mathrm{dx}^\mathrm{i} \mathrm{dx}^\mathrm{j}],
\label{pertmet}
\end{equation}
where $ a(\eta) $ is the scale factor depending on conformal time $\eta $, and $ \psi$ and
$\phi$ are scalar potentials representing gravitational potential and spatial curvature, respectively. The $ (1+2\psi) $ and $(1-2\phi) $ terms modify the temporal and spatial components of the metric, accounting for small perturbations in the spacetime geometry. This gauge simplifies calculations and is commonly used to study linear perturbations.
Furthermore, in the absence of anisotropic stresses, the equations of Einstein's gravity theory require that the metric potentials $\phi$ and $\psi$ are equal. However, this equality does not generally hold in models of modified gravity. 
The perturbed conservation equations, taking into account perturbed metrics and perturbed energy-momentum tensors, yield the following evolution equations for the perturbations\citep{Ma and Bertschinger1995,Marcondes2016,Putter2010}: 
 \begin{align} \notag  
&\dot\delta =-\Big[ 3\mathcal{H}\Big(\frac{\delta p}{\delta\rho}-w_{\mathrm{de}} \Big)-\frac{\bar Q_{0}}{\bar{\rho}}\Big]\delta -\left(1+w_{\mathrm{de}}\right)(\theta -3\dot\phi)\\
&\;\;\;\;\; -\frac{\delta{\bar Q_{0}}}{\bar{\rho}},\label{eq:stbh}\\ \notag
&\dot{\theta}=-\big[\mathcal{H}\left(1-3c^{2}_{\mathrm{a}}\right)-\frac{\bar Q_{0}}{\bar{\rho}}\big]\theta +\frac{\delta p}{\delta\rho}\frac{k^{2}\delta}{\left(1+w_{\mathrm{de}}\right)}+k^{2}\phi \\
&\;\;\;\;\;\; + \frac{i k^{i}\delta \bar Q_{i}}{\bar{\rho}(1+w_{\mathrm{de}})}.  \label{eq5} 
 \end{align}
where dot denotes the derivative with respect to the conformal time, $ \eta $, which is related to the physical time, $t $, through the scale factor, a ($a d\eta=dt$). The variables $k^{i}$, $ \delta\equiv\delta\rho/\rho $, and $ \theta$ represent the components of the wavevector in Fourier space, density contrast, and divergence of the peculiar velocity, respectively. The parameter $ w_{\mathrm{de}}$ corresponds to the EoS of DE, taking different values depending on whether the perturbations are associated with dust($ w_{\mathrm{de}}$=0) or DE. And, $\delta Q_{\mu} $ are the perturbations to the exchange of energy -momentum in the perturbed conservation equations. Lastly, the parameter $ c_{\mathrm{a}}^{2} $ represents the squared adiabatic sound speed of the DE perturbations, and its definition is as follows:
\begin{equation}
c^{2}_{\mathrm{a}}=w_{\mathrm{de}}-\frac{\dot{w}_{\mathrm{de}}}{3\mathcal{H}\left(1+w_{\mathrm{de}}\right)}\label{E:m66}
\end{equation}
To investigate perturbations of DE, it is useful to introduce an effective sound speed, $ c_{\mathrm{eff}}$, specifically for DE perturbations. This quantity is defined as follows\citep{Bean et al.2004}:
\begin{equation}
\frac{\delta p}{\delta\rho}=c^{2}_{\mathrm{eff}} +3\mathcal{H}\left(1+w_{\mathrm{d}}\right)\left(c^{2}_{\mathrm{eff}}-c^{2}_{\mathrm{a}}\right)   
\frac{\theta}{\delta}\frac{1}{k^{2}}\label{E77}
\end{equation}
Additionally, in this context, the Poisson equation can be expressed as follows\citep{Lima1997}:
\begin{equation}\label{eq:pois}
k^{2}\phi = - 4\pi G a^{2}(\delta\rho +3\delta p)
\end{equation}
where $ \delta\rho = \delta\rho_{\mathrm{dm}}+\delta\rho_{\mathrm{de}} $ and  $ \delta p = \delta p_{\mathrm{dm}}+\delta p_{\mathrm{de}} $. After that, using quantities $ \delta p_{\mathrm{dm}}=0 $, $ \delta p_{\mathrm{de}}= c^2_{\mathrm{eff}}\delta\rho_{\mathrm{de}} $, $ \delta\rho_{\mathrm{de}}=\rho_{\mathrm{de}}\delta_{\mathrm{de}} $, $ \delta\rho_{\mathrm{dm}}=\rho_{\mathrm{dm}}\delta_{\mathrm{dm}} $ in   
Eq. (\ref{eq:pois}), the Poisson equation can be written as:
\begin{equation}\label{eq:pois2}
- k^{2}\phi =\frac{3}{2}\mathcal{H}^{2}\big[\Omega_{\mathrm{dm}}\delta_{\mathrm{dm} }+(1+3c_{\mathrm{eff}}^2)\Omega_{\mathrm{de}}\delta_{\mathrm{de}}\big], 
\end{equation} 
where, $\Omega_{\mathrm{dm}}$ and $\Omega_{\mathrm{de}}$ represent the fractional densities of DM and DE respectively. These fractional densities are defined as 
the ratio of the densities $\rho_{\mathrm{dm}}$ and $\rho_{\mathrm{de}}$ to the critical density of the Universe. Also, the $\rho^{0}_{\mathrm{crit}}$, is defined as $\rho^{0}_{\mathrm{crit}}=3H^2_{0}/8\pi G$.
 
In a matter-dominated Universe with a small DE component, the gravitational potential $\phi$ can be approximated as a constant in the linear perturbation regime on sub-horizon scales$(k^{2}\gg\mathcal{H}^{2})$. This assumption is confirmed by the fact that most observed structures, which formed during the matter-dominated era, align with this assumption. This simplification allows for easier analysis of the evolution of perturbations and the growth of structures. However, this assumption is only valid under specific conditions and may not hold in other regimes or on larger scales\citep{Abramo2009}.

To obtain second-order coupled differential equations describing the evolution of  DE and (\ref{eq:stbh} \& \ref{eq5}) as follows:
Firstly, by manipulating Eq. (\ref{E77}), we can obtain the following relation:
\begin{align} \notag
-3\mathcal{H}\frac{\delta p}{\delta \rho}\delta &= -3\mathcal{H}c_{\mathrm{eff}}^{2}\delta -9\frac{\mathcal{H}^{2}}{k^{2}}(1+w_{\mathrm{de}})(c_{\mathrm{eff}}^{2}-c^{2}_{\mathrm{a}})\theta \\
&\simeq -3\mathcal{H}c_{\mathrm{eff}}^{2}\delta \label{Eee}
\end{align}
In the regime of sub-horizon scales ($k^2 \gg \mathcal{H}^2$), we can neglect the second term on the right-hand side of Eq. (\ref{Eee}). 
Secondly, according to Eq.(\ref{E77}), we can express this relation as:
\begin{equation}
k^{2}\frac{\delta p}{\delta \rho}\delta = k^{2}c_{\mathrm{eff}}^{2}\delta + 3\mathcal{H}(1+w_{\mathrm{de}})(c_{\mathrm{eff}}^{2}-c^{2}_{\mathrm{a}})\theta \label{Ebb}
\end{equation}
Now, by substituting Eqs.(\ref{Eee} \& \ref{Ebb}) into Eqs.(\ref{eq:stbh} \& \ref{eq5}), we can express them in the following form:
\begin{align}
&\dot\delta +3\mathcal{H}c^{2}_{\mathrm{eff}}\delta -3\mathcal{H}w_{\mathrm{de}}\delta -\frac{\bar Q_{0} }{\bar{\rho}}\delta +(1+w_{\mathrm{de}})\theta =0 ,\label{E12h}\\
&\dot{\theta}+ \Big[\mathcal{H}\left(1-3c^{2}_{\mathrm{eff}}\right)-\frac{\bar Q_{0}}{\bar{\rho}}\Big]\theta  -k^{2}\phi -
\frac{k^{2}\mathrm{c}^{2}_{\mathrm{eff}}}{1+w_{\mathrm{de}}}\delta  =0 . 
\label{eee} 
\end{align}
Morermore, in order to derive Eqs. (\ref{E12h} \& \ref{eee}) (see also  \citep{Marcondes2016}), we ignore $\delta\bar Q_{\mu}$. 
We remind that Eqs.(\ref{eq:stbh} \& \ref{eq5}) or their equivalent  Eqs.(\ref{E12h} \& \ref{eee}) can be used separately for the components of DE and DM. Based on this, we initially utilize Eqs.(\ref{E12h} \& \ref{eee}) to obtain a second-order equation that describes the evolution of DE perturbations. By eliminating $\theta$ from the system of Eqs.(\ref{E12h} \& \ref{eee}), we can derive following equation for $\delta_{\mathrm{de}}$ in terms of conformal time.
\begin{align}\label{eww}
& \ddot {\delta}_{\mathrm{de}}+ \mathcal{\tilde A}_{\mathrm{de}}\dot {\delta}_{\mathrm{de}}+ \mathcal{\tilde B}_{\mathrm{de}} {\delta}_{\mathrm{de}}= \mathcal{\tilde S}_{\mathrm{de}}
\end{align}
where the coefficients $ \mathcal{\tilde A}_{\mathrm{de}}$, $ \mathcal{\tilde B}_{\mathrm{de}}$, and $ \mathcal{\tilde S}_{\mathrm{de}}$ are defined as follows:
\begin{eqnarray} \notag
\tilde{\mathcal{ A}}_{\mathrm{de}}&=&\mathcal{H}(1-3w_{\mathrm{de}})+3\mathcal{H}(c^{2}_{\mathrm{a}}-w_{\mathrm{de}})-2\frac{\bar Q_{0}}{{\rho}_{\mathrm{de}}}\\ \notag
 \mathcal{\tilde B}_{\mathrm{de}}&=&3\mathcal{H}^{2}(c^{2}_{\mathrm{eff}}-w_{\mathrm{de}})\Big[1+\frac{\mathcal{\dot H}}{\mathcal{H}^{2}}-3(w_{\mathrm{de}}+c^{2}_{\mathrm{eff}}-c^{2}_{\mathrm{a}})\Big]\\ \notag
&+&k^{2}c^{2}_{\mathrm{eff}}-3\mathcal{H}\dot{w}_{\mathrm{de}}
+\mathcal{H}\Big[3\big (2w_{\mathrm{de}}-c^{2}_{\mathrm{a}}\big)-1\Big]\frac{\bar Q_{0}}{{\rho_{\mathrm{de}}}}\\  \notag 
&+&\Big(\frac{\bar Q_{0}}{\rho_{\mathrm{de}}}\Big)^{2}  
 -\frac{d}{d\eta}\Big(\frac{\bar Q_{0}}{\rho_{\mathrm{de}}}\Big)\\
 \mathcal{\tilde S}_{\mathrm{de}}&=&-(1+w_{\mathrm{de}})k^{2}\phi  \label{err}
\end{eqnarray}
where $- k^{2}\phi $ is expressed by Poisson Eq. (\ref{eq:pois2}).
Likewise, by utilizing Eqs. (\ref{E12h} \& \ref{eee}), we can derive a second-order equation that describes the evolution of DM perturbations. In this case, we set \(w_{\mathrm{d}} = c^{2}_{\mathrm{eff}} = c^{2}_{\mathrm{a}} = 0\). The resulting equation is obtained as follows:
 \begin{align}\label{eqq}
& \ddot {\delta}_{\mathrm{dm}}+ \mathcal{\tilde A}_{\mathrm{dm}}\dot {\delta}_{\mathrm{dm}}+ \mathcal{\tilde B}_{\mathrm{dm}} {\delta}_{\mathrm{dm}}= \mathcal{\tilde S}_{\mathrm{dm}}
\end{align}
where, in this case, the coefficients $   \mathcal{\tilde A}_{\mathrm{dm}}$, $   \mathcal{\tilde B}_{\mathrm{dm}}$, and $   \mathcal{\tilde S}_{\mathrm{dm}}$ are defined as follows:
\begin{align} \notag
& \mathcal{\tilde A}_{\mathrm{dm}}=\mathcal{H}-2\frac{\bar Q_{0}}{{\rho}_{\mathrm{dm}}}\\ \notag
& \mathcal{\tilde B}_{\mathrm{dm}}=  -\mathcal{H}\frac{\bar Q_{0}}{{\rho_{\mathrm{dm}}}} +\Big(\frac{\bar Q_{0}}{\rho_{\mathrm{dm}}}\Big)^{2}
 -\frac{d}{d\eta}\Big(\frac{\bar Q_{0}}{\rho_{\mathrm{dm}}}\Big)\\ 
& \mathcal{\tilde S}_{\mathrm{dm}}=-k^{2}\phi  \label{euuu}
\end{align}
Additionally, by utilizing the expressions $\frac{d}{d\eta}=a \mathcal{H}\frac{d}{da}$ and $\frac{d^{2}}{d\eta^{2}}=(a\mathcal{H}^{2}+a\mathcal{\dot{H}}) \frac{d}{da} +a^{2}\mathcal{H}^{2}
\frac{d^{2}}{da^{2}}$, along with Eq. (\ref{eq:pois2}), one can represent Eqs. (\ref{eww} \& \ref{eqq}) in terms of the scale factor. Thus, we obtain the following equations:
\begin{align}\label{ewq}
& {\delta}^{ \prime \prime}_{\mathrm{de}}+A_{\mathrm{de}} {\delta}^{ \prime}_{\mathrm{de}}+B_{\mathrm{de}} {\delta}_{\mathrm{de}}=S_{\mathrm{de}}\\
&  {\delta}^{\prime\prime}_{\mathrm{dm}}+{A}_{\mathrm{dm}} {\delta}^{\prime}_{\mathrm{dm}}+ {B}_{\mathrm{dm}} {\delta}_{\mathrm{dm}}= {S}_{\mathrm{dm}} \label{ecq}
\end{align}
where, the prime denotes the derivative with respect to the scale factor. The coefficients $ {A}_{\mathrm{de}} $, $ {B}_{\mathrm{de}} $, and $ {S}_{\mathrm{de}} $ are defined as follows:
\begin{eqnarray} \notag
{A}_{\mathrm{de}}&=&\frac{3}{a}+\frac{H^{\prime}}{H}+\frac{3}{a}(c^{2}_{\mathrm{a}}-2w_{\mathrm{de}})-\frac{2}{a^{2}H}\frac{\bar Q_{0}}{{\rho}_{\mathrm{de}}}\\ \notag
{B}_{\mathrm{de}}&=&\dfrac{3}{a}(c^{2}_{\mathrm{eff}}-w_{\mathrm{de}})\Big[\frac{2}{a}+\frac{ H^{\prime}}{H}-\frac{3}{a}(w_{\mathrm{de}}+c^{2}_{\mathrm{eff}}-c^{2}_{\mathrm{a}})\Big]\\ \notag
&+&\frac{k^{2}c^{2}_{\mathrm{eff}}}{a^{4}H^{2}}-\frac{3}{a}w^{\prime}_{\mathrm{de}}+\frac{1}{a^{3}H^{2}}\Big[3\big (2w_{\mathrm{de}}-c^{2}_{\mathrm{a}}\big)-1\Big]\frac{\bar Q_{0}}{{\rho_{\mathrm{de}}}} \\  
&+&\frac{1}{a^{4}H^{2}}\Big(\frac{\bar Q_{0}}{\rho_{\mathrm{de}}}\Big)^{2}
 -\frac{1}{a^{2}H}\frac{d}{da}\Big(\frac{\bar Q_{0}}{\rho_{\mathrm{de}}}\Big)\\ \notag
{S}_{\mathrm{de}}&=&\frac{3}{2a^{2}} \left(1+w_{\mathrm{de}}\right) \Big[\Omega_{\mathrm{dm}}\delta_{\mathrm{dm}}+
\Omega_{\mathrm{de}}\delta_{\mathrm{de}}\left(1+3c_{\mathrm{eff}}^{2}\right)\Big]\notag   \label{err}
\end{eqnarray}
In addition, the coefficients $ {A}_{\mathrm{dm}} $, $ {B}_{\mathrm{dm}} $, and $ {S}_{\mathrm{dm}} $ can be defined as follows:
%%%%%%%%%%%%%%%%%%%%%%%%%%%%%%
\begin{align} \notag
&{A}_{\mathrm{dm}}=\frac{3}{a}+\frac{H^{\prime}}{H}-\frac{2}{a^{2}H}\frac{\bar Q_{0}}{{\rho}_{\mathrm{dm}}}\\ \notag
&{B}_{\mathrm{dm}}=-\frac{1}{a^{3}H^{2}}\frac{\bar Q_{0}}{{\rho_{\mathrm{dm}}}}  
+\frac{1}{a^{4}H^{2}}\Big(\frac{\bar Q_{0}}{\rho_{\mathrm{dm}}}\Big)^{2}
 -\frac{1}{a^{2}H}\frac{d}{da}\Big(\frac{\bar Q_{0}}{\rho_{\mathrm{dm}}}\Big)\\ 
&{S}_{\mathrm{dm}}=\frac{3}{2a^{2}}\Big[\Omega_{\mathrm{dm}}\delta_{\mathrm{dm}}+\Omega_{\mathrm{de}}\delta_{\mathrm{de}}\Big]    \label{eroo}
\end{align}
Where $ \bar Q_{0} $, $\frac{\bar Q_{0}}{\rho_{\mathrm{dm}}}  $, and $  \frac{\bar Q_{0}}{\rho_{\mathrm{de}}}$ for both models IDE I and IDE II are summarized in Table \ref{tab1222}.    
By solving the the coupled Eqs. (\ref{ewq} and \ref{ecq}) numerically from an initial scale factor of $a_{\mathrm{i}}=10^{-3}$ to the current time ($a=1$), we can obtain the density contrasts of the $ \delta_{\rm dm} $ and $\delta_{\rm de}$.

 The effect of clustered and non-clustered DE on DM perturbations can be explored by considering the effective sound speed parameter $\mathrm{c}_{\mathrm{eff}}$, where $\mathrm{c}_{\mathrm{eff}}\simeq 0$ for clustered DE and $\mathrm{c}_{\mathrm{eff}}\simeq 1$ for non-clustered or homogeneous DE. Also, in the case of non-clustered DE, we can simplify the equations by setting $\delta_{\rm de}=0$.
This allows us to determine the evolution of the density contrasts $ \delta_{\rm dm} $ and $\delta_{\rm de}$ as a functions of the scale factor via numerical integration with following appropriate initial conditions
\citep{Batista2013, Abramo2009}.
\begin{align}
&\delta_{\mathrm{dm,i}}=-2\phi_{i}\Big(1+\frac{k^{2}}{3{\mathcal{H}_{i}}^{2}}\Big) \;\;;\;\ \delta^{\prime}_{\mathrm{dm,i}}=-\frac{2}{3}\frac{k^{2}}{\mathcal{H}^{2}_{i}}\phi_{i}\\ \notag &\delta_{\mathrm{de,i}}=(1+w_{\mathrm{di}})\delta_{\mathrm{dm,i}}\\
&\delta_{\mathrm{de,i}}^{\prime}=(1+w_{\mathrm{di}})\delta^{\prime}_{\mathrm{dm,i}}+w^{\prime}_{\mathrm{di}}\delta_{\mathrm{dm,i}}
\end{align}
Where $ w_{\rm di}$ means the value of $ w_{\rm de}$ at $ a_{\rm i}$. 
The choice of $k\mathrm{ = 0.1 h Mpc^{-1}}$ ensures that the analysis remains in the linear regime because it falls within the range of scales where the linear approximation is valid. This choice is supported by the assumption that the shape of the power spectrum recovered from galaxy surveys matches the linear matter power spectrum shape for scales $k  \rm{\leq 0.15 h  Mpc^{-1}}$.
Additionally, it is consistent with the power-spectrum normalization $ \sigma_{8}$, which corresponds to $ k\rm{= 0.125 h Mpc^{-1}} $.
The specific value chosen for $ \phi_{i} $, such as  $ \phi_{i} =- 2 \times 10^{-6} $, corresponds to $ \delta _{\rm dm}= 0.08 $ at the present time for $ k\mathrm{ = 0.11 h Mpc^{-1}} $. 
Therefore, the choice of $k\mathrm{= 0.1 h Mpc^{-1}}$ allows for a reliable examination of the growth rate of clustering in the linear regime\citep{Smith2003,Tegmark2004,Percival2007}. In the following section, we will utilize the numerical results derived from solving Eqs. (\ref{ewq} \& \ref{ecq}) to examine the growth rate related to DM.
\begin{table}
 \centering
 \caption[fs8]{The  $f\sigma_{8}(z) $ data points and their References.}
\begin{tabular}{lcc|cccc} 
\hline
\hline
  $z_{i}$ &  $f\sigma_{8}(z_{i}) $& Refs.& $ z $ &  $f\sigma_{8}(z) $& Refs.\\ 
\hline\hline
 0.02& $ 0.428 \pm0.0465$ & \citep{Huterer et al.:217}&0.3& $ 0.407 \pm0.055$ & \citep{Rita2012}\\
 
0.02& $ 0.398 \pm0.065$& \citep{Hudson:Turnbull2013}& 0.02& $ 0.314 \pm0.048$&\citep{Hudson:Turnbull2013}&\\

 0.38& $ 0.477 \pm0.051$&\citep{Florian2017}& 0.51& $ 0.453 \pm0.050$&\citep{Florian2017}&\\

 0.61& $ 0.410 \pm0.044$&\citep{Florian2017}& 0.76& $ 0.440 \pm0.040$ &\citep{Michael2016}&\\

 1.05& $ 0.280 \pm0.080$&\citep{Michael2016}& 0.32& $ 0.427 \pm0.056$ &\citep{Hector2017}&\\

 0.57& $ 0.426 \pm0.029$&\citep{Hector2017}&0.38& $ 0.497 \pm0.045$ &\citep{Shadab2017}&\\

0.51& $ 0.458 \pm0.038$&\citep{Shadab2017}&0.61& $ 0.436 \pm0.034$ &\citep{Shadab2017}&\\

0.31& $ 0.469 \pm0.098$&\citep{Yuting2017}&0.36& $ 0.474 \pm0.097$&\citep{Yuting2017}&\\

0.40& $ 0.473 \pm0.086$&\citep{Yuting2017}&0.44& $ 0.481 \pm0.076$&\citep{Yuting2017}&\\

0.52& $ 0.488 \pm0.065$&\citep{Yuting2017}&0.48& $ 0.482\pm0.067$&\citep{Yuting2017}&\\

0.59& $ 0.481\pm0.066$&\citep{Yuting2017}&0.56& $ 0.482\pm0.067$&\citep{Yuting2017}&\\

0.64& $ 0.486\pm0.070$&\citep{Yuting2017}&0.10 &$ 0.370 \pm0.130$ &\citep{Feix et al2015}&\\
 
 0.15 &$ 0.490 \pm0.145$&\citep{Howlettetal2015}&0.17 &$ 0.510 \pm0.060$&\citep{SongPercival2009}&\\

 0.18 & $ 0.360 \pm0.090$ &\citep{Blake et al:2013}&0.38 & $ 0.440 \pm0.060$&\citep{Blake et al:2013}&\\
 
0.25 & $ 0.3512 \pm0.0583$ &\citep{Samushia et al. 2012}&0.37 & $ 0.4602 \pm0.0378$&\citep{Samushia et al. 2012}&\\

 0.32 & $ 0.384 \pm0.095$&\citep{Sanchez et al. 2014}& 0.59 &$ 0.488 \pm0.060$ &\citep{Chuang et al. 2016}&\\
 
 0.44& $ 0.413 \pm0.080$&\citep{Blake et al. 2012}& 0.60& $ 0.390 \pm0.063$& \citep{Blake et al. 2012}&\\

 0.73 & $ 0.437 \pm0.072$&\citep{Blake et al. 2012}&0.60 & $ 0.550 \pm0.120$ &\citep{Pezzotta et al. 2017}&\\
 
 1.52 & $ 0.420 \pm0.076$&\citep{Hector2018}& 0.50 & $ 0.427 \pm0.043$&\citep{Rita2012}&\\
   
 0.86 & $ 0.400 \pm0.110$&\citep{Pezzotta et al. 2017}&1.40 & $ 0.482 \pm0.116$ &\citep{Okumura et al. 2016} &\\

0.978 & $ 0.379 \pm0.176$ &\citep{Zhao et al2019} &1.23 & $ 0.385 \pm0.099$ &\citep{Zhao et al2019} &\\   

1.526  & $ 0.342 \pm0.070$ &\citep{Zhao et al2019} &1.944  & $0.364 \pm0.106$ &\citep{Zhao et al2019} &\\ 

\hline\hline
 \label{tab1}
 \end{tabular}
 \begin{flushleft}
  \vspace{-0.5cm}
  {\small}
 \end{flushleft}
\end{table}
\subsection{\textbf{Growth of matter perturbations}} \label{growth}
By numerically solving Eqs.(\ref{ewq} \& \ref{ecq}), we can determine the theoretical prediction for the quantity $f\sigma_{8}$. 
The quantity $f$ represents the linear growth rate of matter perturbations as a function of redshift (z). It quantifies how structures form and evolve, and is defined as follows\citep{Nesseris2017}:
\begin{equation}\label{ffff2}
f=\dfrac{d\ln\delta_{dm}}{d\ln a}=-\frac{1+z}{z}\frac{d\ln \delta_{m}}{d\ln z}
\end{equation}
On the other hand, $\sigma_{8}(z)$ quantifies the growth of root-mean-square mass fluctuations in spheres with radius $8 \mathrm{Mpch^{-1}}$\citep{Nesseris2008}, and can be calculated in the linear regime as $\sigma_{8}(z)=\sigma_{8,0}\frac{\delta_{m}(z)}{\delta_{m}(z=0)}$. Also, $\sigma_{8}(z)$ characterizes the level of clustering or fluctuations in the distribution of matter on large scales. 
Furthermore, we can rescale the parameter $\sigma_{8,0}$ as $\sigma_{8,0}=\frac{\delta_{m}(z=0)}{\delta_{m,\Lambda}(z=0)}\sigma_{8,\Lambda}$ to obtain appropriate parameters for evaluating different cosmological models, particularly in the context of IDE models. The $f(z)\sigma_{8}(z)$ measurement provides insights into the perturbations of the galaxy density, represented as $\delta_{g}$, which is related to the perturbations in DM through the bias factor $ b$, defined as $ b=\delta_{g}/ \delta_{m}$\citep{Michael2016}.  
The independence of $f(z)\sigma_{8}(z)$ from the bias factor, as shown by Song and Percival\citep{SongPercival2009}, is significant because it allows for more reliable and robust discrimination between different IDE models based on this quantity.
In conclusion, the validity of various IDE models can be assessed by comparing the theoretical predictions of $f\sigma_{8}(z)$ with observational data. This is accomplished by calculating the $\chi^{2}_{\mathrm{growth}}$ statistic, which can be expressed as follows:
\begin{equation}
\chi^{2}_{\mathrm{growth}} = \sum_{i=1}^{44} \frac{\left[(f\sigma_{8})_{\mathrm{th}}(z_i) - (f\sigma_{8})_{\mathrm{obs}}(z_i)\right]^2}{\sigma_{\mathrm{obs}}^2(z_i)}
\end{equation}
where, $(f\sigma_{8})_{\mathrm{th}}(z_i)$ represents the theoretical prediction at the redshift $z_{i}$, while $(f\sigma_{8})_{\mathrm{obs}}(z_i)$ and $\sigma_{\mathrm{obs}}(z_i)$ denote the measured values and uncertainties, respectively.
The dataset used in this study, consisting of 44 measurements of $f\sigma_{8}(z)$, is displayed in Table \ref{tab1}.

\begin{figure}
\centering
\setlength\fboxsep{0pt}
\setlength\fboxrule{0pt}
\fbox
{\includegraphics[width=8.5cm]{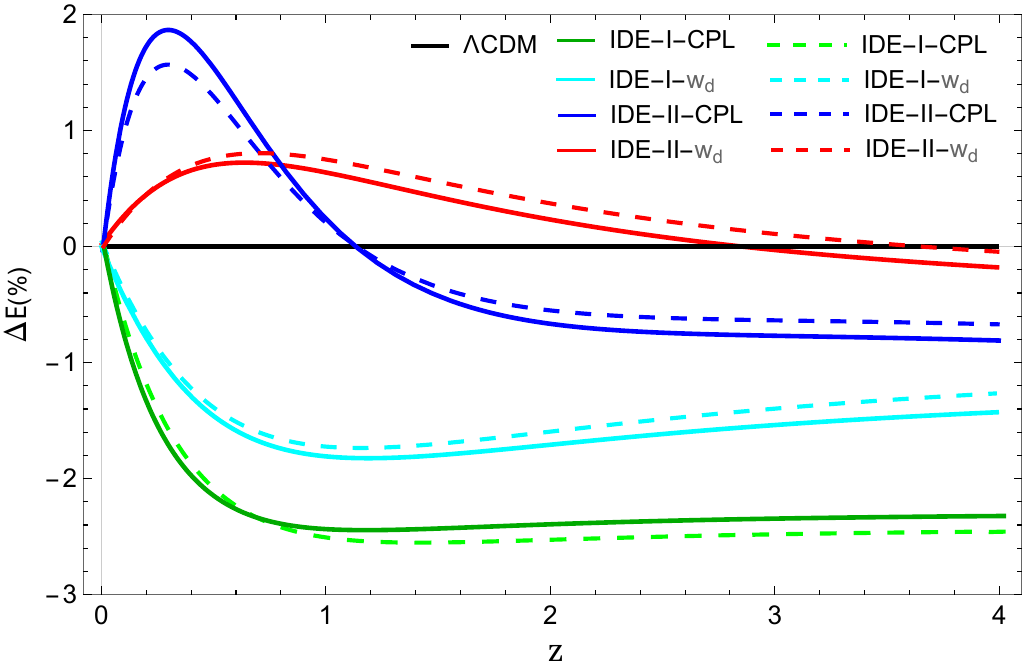}}
{\includegraphics[width=8.5cm]{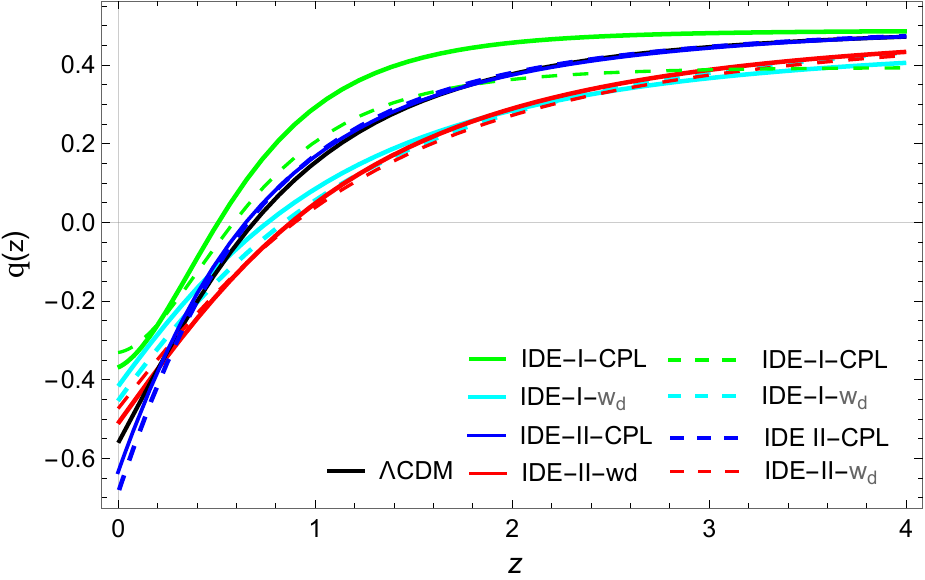}}
{\includegraphics[width=8.5cm]{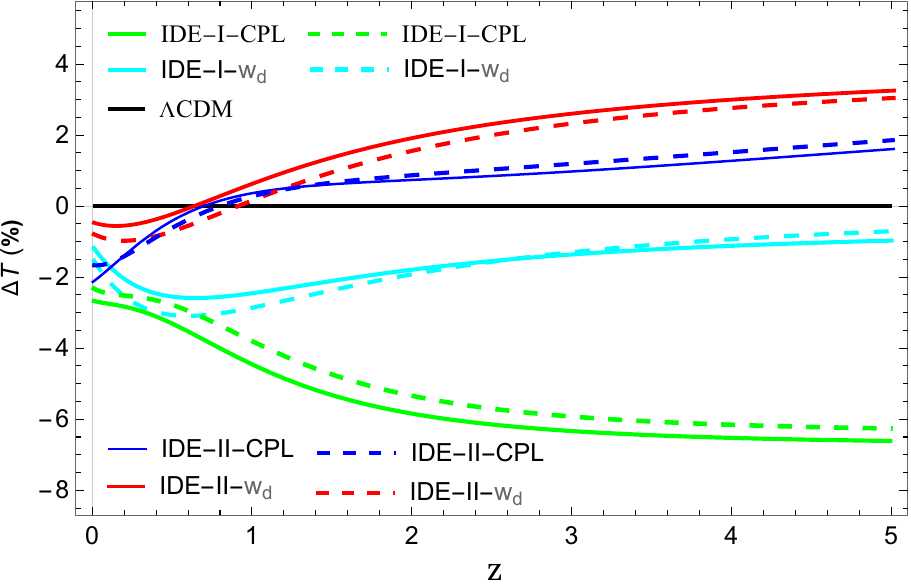}}
\caption{Upper panel: The $\Delta E(\%) $ of the IDE models compared to the $ \Lambda$CDM model (see Eq. \ref{deltah11}). 
Middle panel: The evolution of the deceleration parameter for various
models. Lower panel: The $\Delta T(\%) $ of the IDE models compared to its value in the standard $\Lambda$CDM model(see Eq. \ref{delTT}), using the best-fit values listed in Tab.\ref{tabhom} for the IDE models. The various IDE models have been specified by different colors and line styles in the inner panels of the figure. The dashed (solid) line represents the homogeneous (clustered) case of DE.}
\label{f7888}
\end{figure}
\section{\textbf{IDE MODELS VERSUS DATA ANALYSIS }}\label{sec5}
In this section, we will examine the IDE models considered in this study by following a two-step approach.

\begin{figure*}
\centering
\setlength\fboxsep{0pt}
\setlength\fboxrule{0pt}
\fbox
{\includegraphics[width=8.6cm]{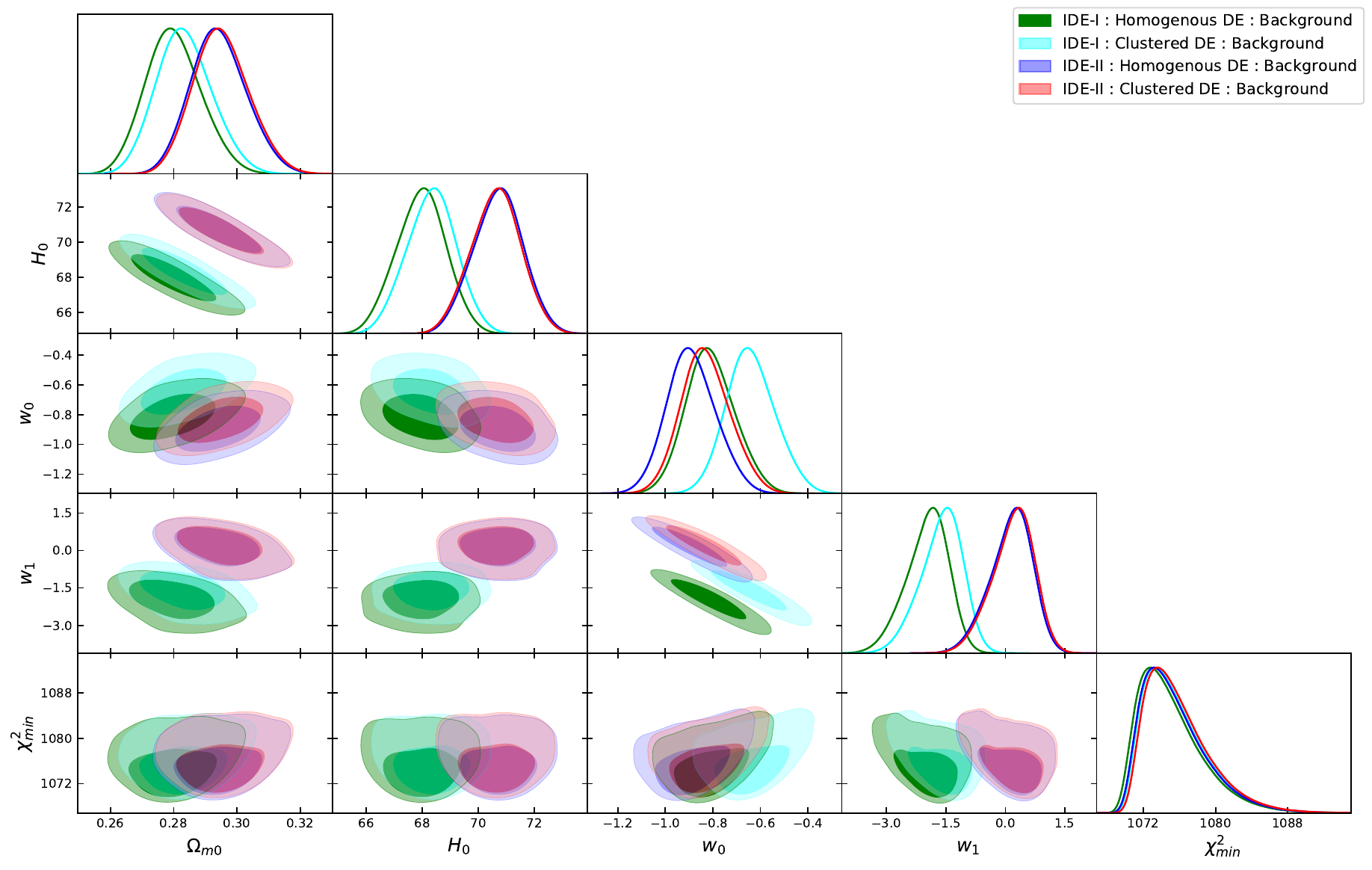}}
{\includegraphics[width=8.6cm]{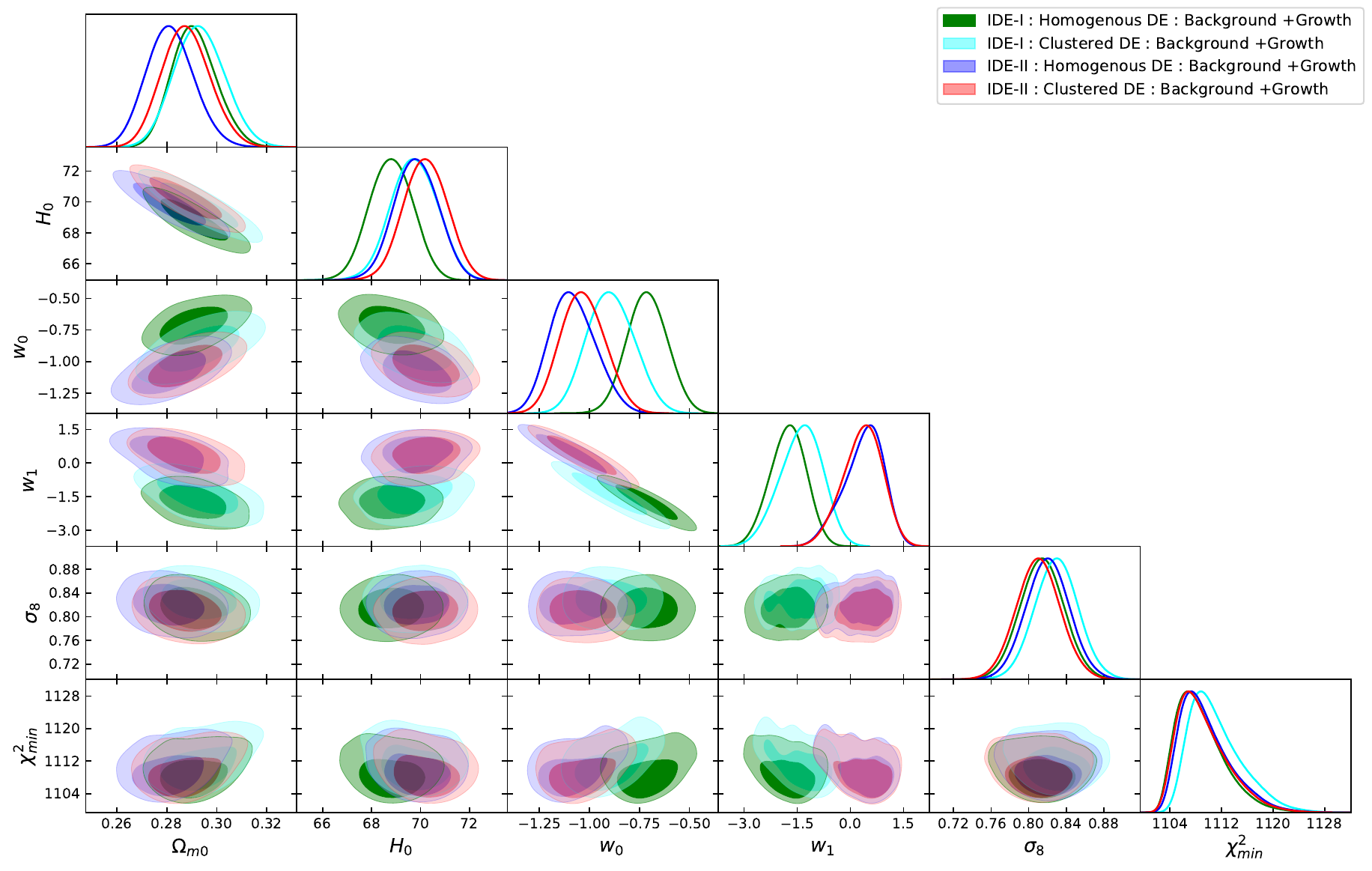}}
{\includegraphics[width=8.6cm]{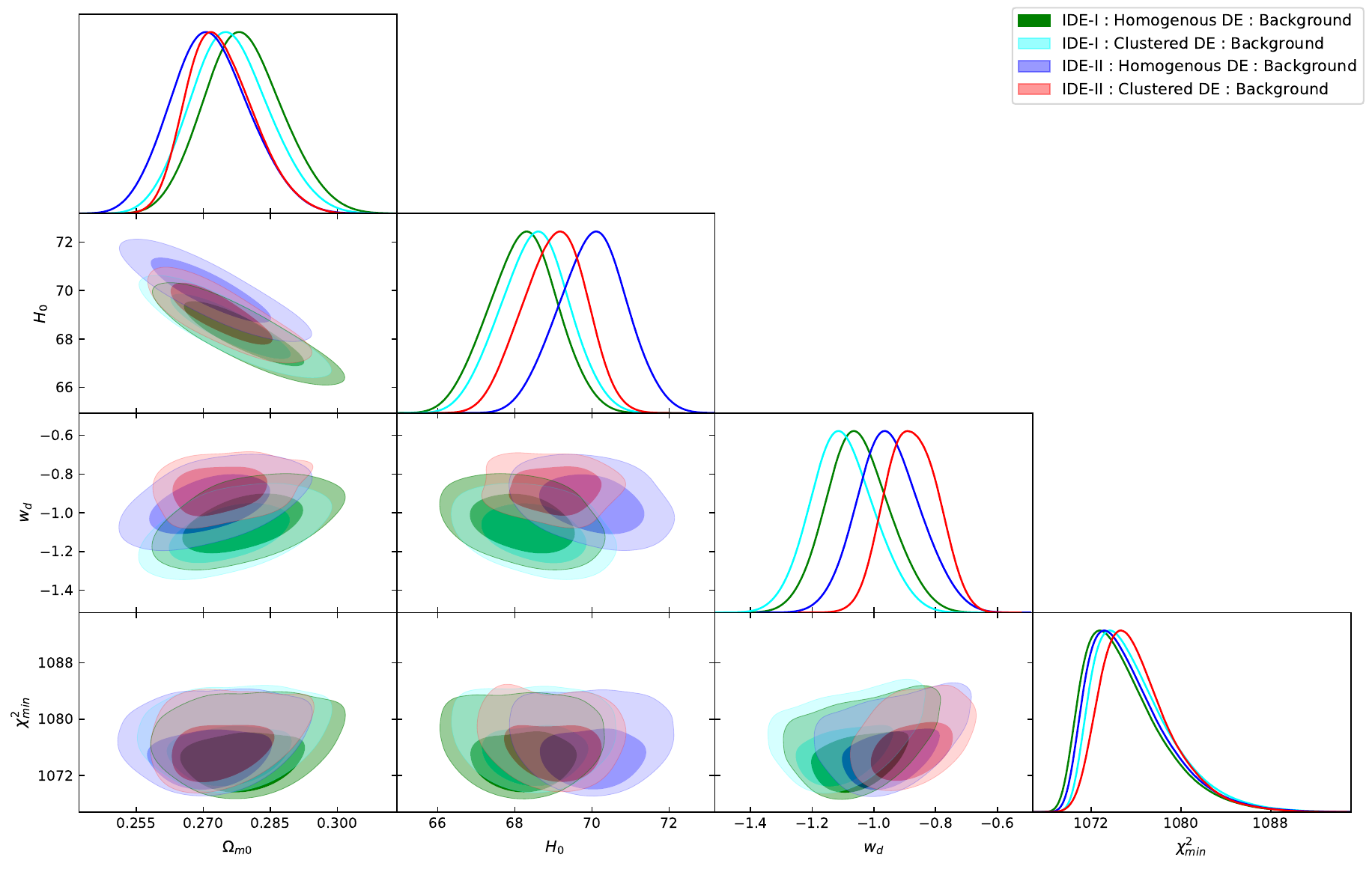}}
{\includegraphics[width=8.6cm]{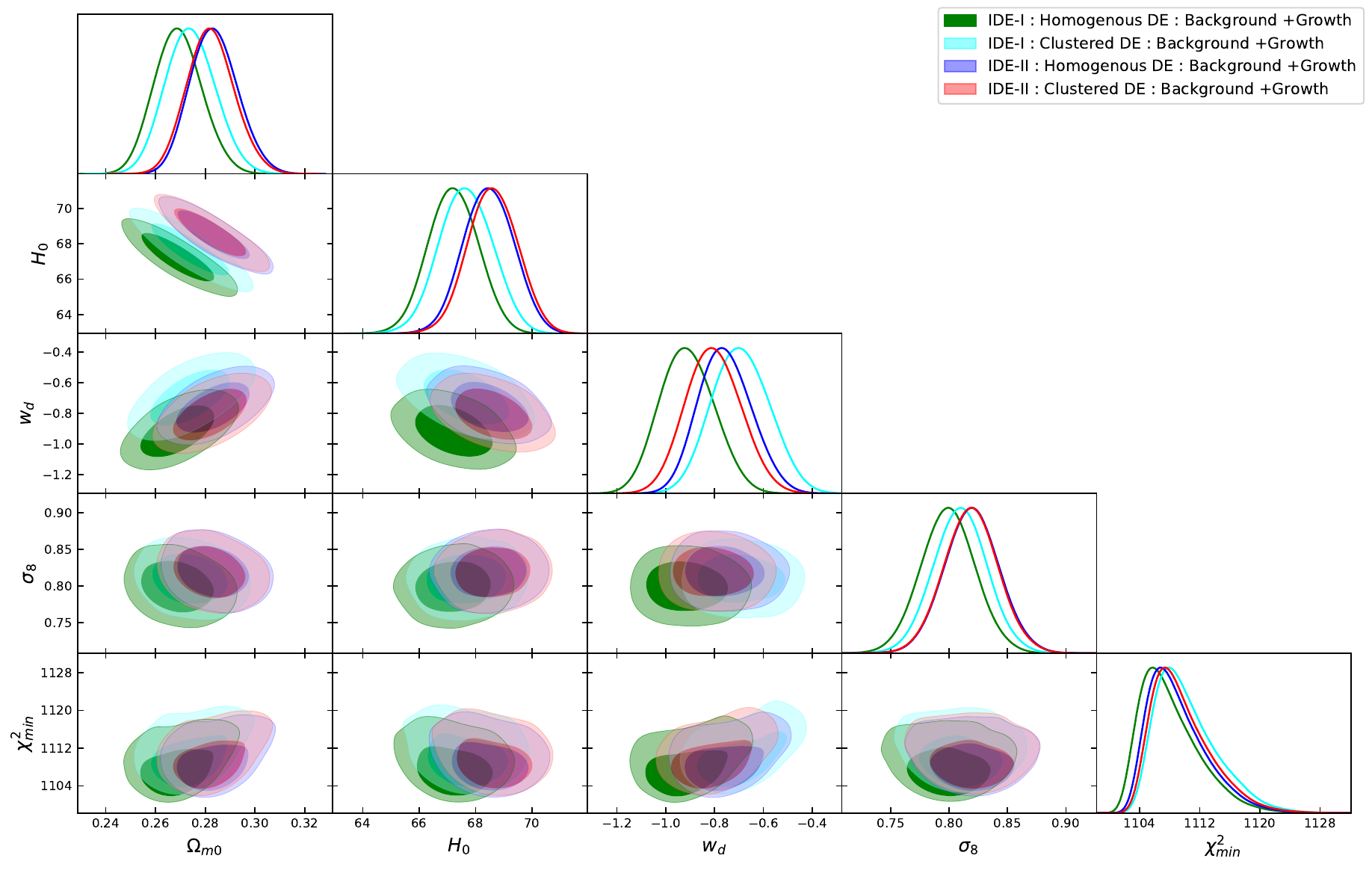}}
\caption{The confidence levels for $1\sigma$ and $2\sigma$ constraints on the IDE-I and IDE-2 models using different datasets. The upper left panel shows confidence levels obtained solely from background data, while the upper right panel displays constraints obtained by combining background and growth rate data. These upper panel constraints are presented for both homogeneous and clustered cases of DE, assuming the equation of state (EoS) parameter of DE follows the CPL parameterization. Similarly, the lower panels (left and right) depict constraints when the EoS parameter of DE, $ w_{\mathrm{d}} $, is considered constant.}
\label{f732q}
\end{figure*}

Initially, we perform an $\mathrm{MCMC}$ analysis to constrain the free parameters of the models based on the latest available background data (see Sec. \ref{sec3} and Eq. (\ref{xi22})). Subsequently, we provide a concise overview of our data analysis results pertaining to the IDE models, which can be found in Table \ref{tabhom}. Furthermore, left panels of Fig. \ref{f732q} illustrates the confidence levels for $1\sigma $ and $2\sigma $ constraints on the IDE models based on the background datasets. These triangular plots are particularly valuable as they visually indicate the correlations between each pair of free parameters in the models.

The Hubble parameter plays an important role in characterizing the background evolution of the Universe. Moreover, how the Hubble parameter evolves can influence the growth of matter perturbations. Therefore, it is very important to investigate the behavior of the Hubble parameter in the context of IDE models. In light of this, the  upper panel of Fig. \ref{f7888} illustrates the evolution of the percentage deviation of the normalized Hubble parameter E(z) of the models in comparison to the standard $\Lambda \mathrm{CDM}$ model. In other words, it shows the relative deviation of the normalized Hubble parameter of the models from the concordance $\Lambda \mathrm{CDM}$ model, i.e.
\begin{equation}\label{deltah11}
\Delta E(\%)= 100\times\Big[\frac{E(z)_{\mathrm{model}}}{E(z)_{\Lambda \mathrm{CDM}}}-1\Big]
\end{equation}
In the top panel of Fig.\ref{f7888}, it is obvious that the value of quantity $\Delta E(\%) $  associated with IDE-I model, considering the CPL parameterization and a constant value for $ w_{\mathrm{de}} $, exhibits negative values in comparison to $\Lambda \mathrm{CDM}$ model for all z. This finding holds for both the scenarios of homogeneous and clustered DE.
Moreover, in the case of the IDE-II model assuming the CPL parameterization and a constant value for $ w_{\mathrm{de}} $ of  DE, the $\Delta E(\%) $ is positive at ${z\lesssim 1.14}$ and ${z\lesssim 2.93}$, respectively. This is true for both homogeneous and clustered DE. Being positive (negative) value of quantity $\Delta \mathrm{E}(\%) $ relative to the $\Lambda \mathrm{CDM}$ model means that the cosmic expansion in the corresponding IDE model is larger (smaller) compared to the $\Lambda \mathrm{CDM}$ model. 
Moreover, in the right panel of Fig. \ref{ffr7}, we present a comparison between the theoretical evolution of the Hubble parameter, $\mathrm{H(z)} $, and a set of 36 cosmic chronometer data points listed in Table \ref{tabHdata}.

Here, we explore the deceleration parameter, which can be utilized for evaluating IDE models. This parameter is defined as follows:
\begin{equation}\label{q2zzz}
q(z)=-\frac{\ddot{a}}{aH^{2}}=\frac{1}{H(z)}\frac{dH(z)}{dz}(1+z)-1
\end{equation}
\begin{figure}
\centering
\setlength\fboxsep{0pt}
\setlength\fboxrule{0pt}
\fbox
{\includegraphics[width=8.5cm]{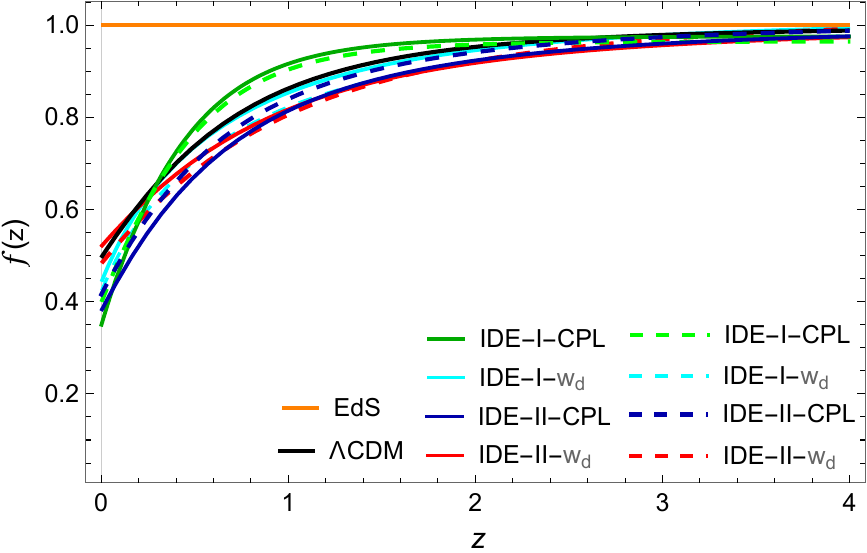}}
{\includegraphics[width=8.5cm]{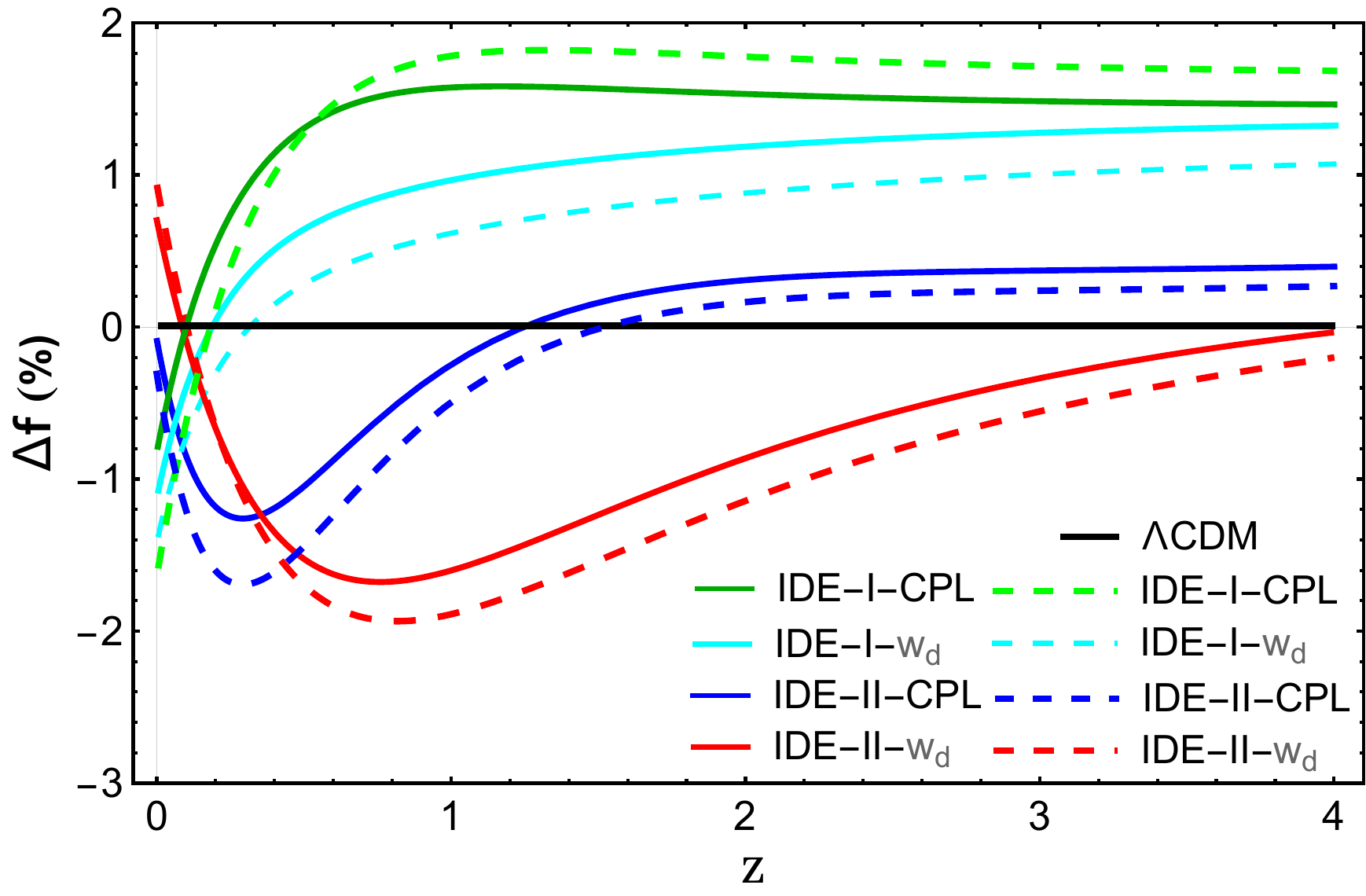}}
\caption{Upper panel: Evolution of the linear growth rate of
matter perturbations (see Eq. \ref{ffff2}) for the various IDE models in terms of the redshift z. Lower panel: The $\Delta f(\%) $ of models compared to standard $ \Lambda \mathrm{CDM} $ model as a function of redshift z (see Eq. \ref{deem}). The color lines and styles utilized in the inner panels of the figure resemble those depicted in Fig.\ref{f7888}}
\label{f733}
\end{figure}
By utilizing Equation (\ref{q2zzz}), we can calculate the transition time, denoted as $z_{t}$, when the Universe undergoes a shift from a decelerated expansion phase ($q > 0$) to an accelerated expansion phase ($q < 0$). This transition time is determined by setting either $q = 0$ or $\ddot{a} = 0$. The middle panel of Fig. \ref{f7888}, illustrates the evolution of the deceleration parameter for the IDE models as a function of the redshift $z$. 
 The values of the transition redshift, $z_{t}$, pertaining to the IDE-I and IDE-II models, considering the CPL parameterization and a constant value for $w_{\mathrm{de}}$, are as follows: 
{\begin{equation}
z_{t}=
\begin{cases}
 \text{{\small Homogenous DE}}\;\;\;\; \text{{\small Clustered DE}}\;\;\;\;\;\;\;\;\text{{\small Model}}\\
\;\;\; \;\;\sim  0.592\; ; \;\;\;\;\;\;\;\;\;\; \;\; \sim 0.506\;\;\;\;\;\;\;\; \;\; \text{{\small IDE-I, CPL;}}\\
\;\;\; \;\;\sim  0.838\; ; \;\;\;\;\;\;\;\;\;\; \;\; \sim 0.736\;\;\;\;\;\;\;\; \;\; \text{{\small IDE-I, }$w_{\mathrm{de}}$;}\\
\;\;\; \;\;\sim  0.643\; ;\;\;\;\;\;\;\;\;\;\; \;\; \sim 0.674\;\;\;\;\;\;\;\; \;\; \text{{\small IDE-II, CPL;}}\\
\;\;\; \;\;\sim  0.791\; ; \;\;\;\;\;\;\;\;\;\; \;\;\sim 0.780\;\;\;\;\;\;\;\; \;\; \text{{\small IDE-II, }$w_{\mathrm{de}}$}
\end{cases} \notag
\end{equation}}
Moreover, the transition redshift, $ z_{\mathrm{t}} $, for $ \Lambda \mathrm{CDM}$ model, is $ z_{t}= 0.692 $. 
It is evident that during the early times, when matter was the dominant component in the Universe, the quantity $ q $ approaches a value of $\frac{1}{2}$. Hence, during the early matter-dominated era, the value of $ q $ indicates a decelerating but slowing expansion.  These outcomes are consistent with the findings reported in the study by Farooq et al. in \citep{Farooq2017}.

The age of the Universe can be used as another parameter for assessing and comparing different models of IDE. By utilizing the best-fitting values provided in Table \ref{tabhom} and applying the following equation, we can compute the age of the Universe.
\begin{equation}\label{tu}
t_{_{U}}=\frac{1}{H_{0}}\int_{0}^{\infty}\frac{dz}{(1+z)E(z)}
\end{equation}
which $ E(z) $ is given by Eq. (\ref{etqq}). The age of the Universe, determined by the Eq.  (\ref{tu}), yields the following results for the IDE models analyzed in this study.
The $  t_{_{U}} $ is computed for both homogeneous and clustered DE scenarios. In the case of the homogeneous DE, $  t_{_{U}} $ for IDE-I(CPL, $ w_{\mathrm{de}}$) = $(13.33, 13.43)\mathrm{Gyr}$ and for IDE-II (CPL, $ w_{\mathrm{de}}$) =$(13.41, 13.53) \mathrm{Gyr}$. Similarly, in the case of clustered DE, the $  t_{_{U}} $ for IDE-I (CPL, $ w_{\mathrm{de}}$) =$(13.28, 13.48)\mathrm{Gyr} $ and for IDE-II (CPL, $ w_{\mathrm{de}}$) = $(13.35, 13.57)\mathrm{Gyr}$. In addition, we indicated that the value of $t_{_{U}}$ for the $\Lambda \mathrm{CDM}$ model is $13.642 \; \mathrm{Gyr}$.
It is worth noting that the age of the Universe, as determined by the Planck (2018) results, is $13.78\;\mathrm{Gyr}$ \citep{Aghanim et al2020}.

Additionally, the lower panel in Fig. \ref{f7888} illustrates the percentage of the relative deviation in the age of the Universe for the IDE models compared to the standard $\Lambda$CDM model. This quantity is defined as follows:
\begin{equation}\label{delTT}
\Delta T(\%)=100\times\Big[\frac{(t_{_{U}})_{\mathrm{model}}}{(t_{_{U}})_{\Lambda \mathrm{CDM}}}-1\Big]
\end{equation}
In the lower panel of Fig. \ref{f7888}, we see that the results of our analysis for the IDE models investigated in this study, the values of $\Delta T(\%)$, are as follows:
\begin{figure*}
\centering
\setlength\fboxsep{0pt}
\setlength\fboxrule{0pt}
\fbox
{\includegraphics[width=8.6cm]{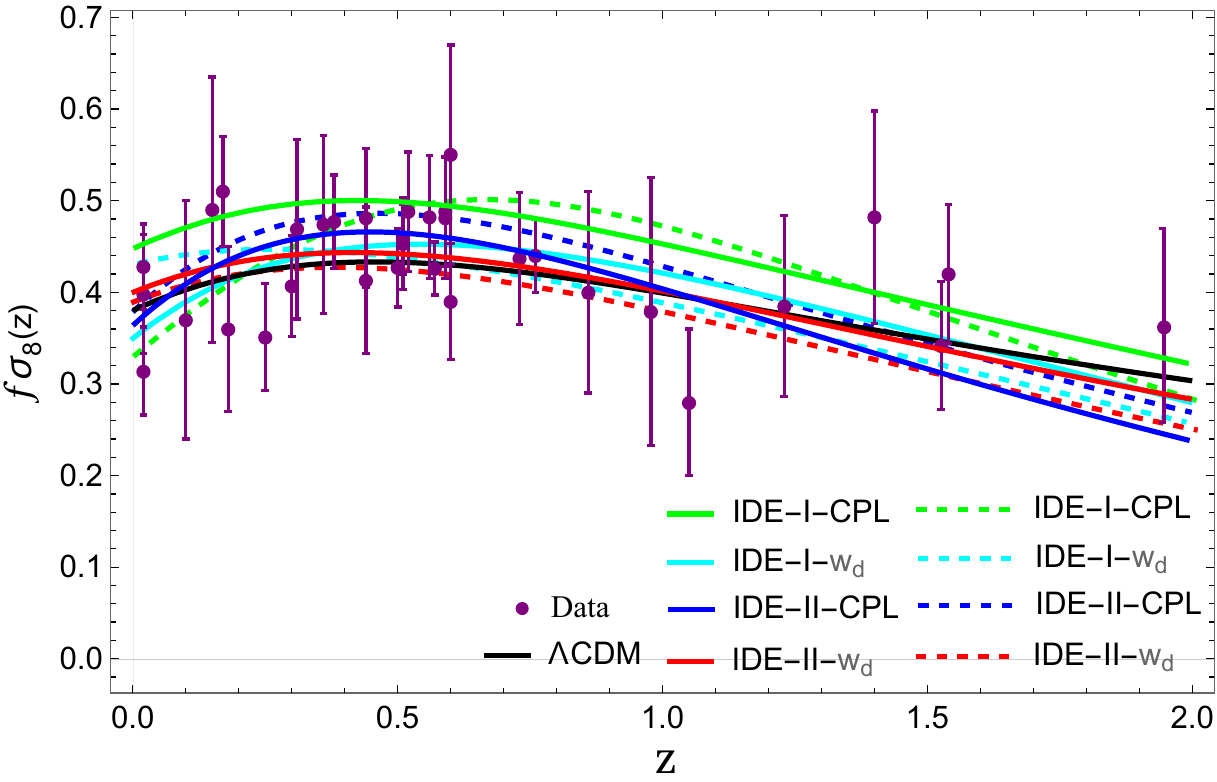}}
{\includegraphics[width=8.6cm]{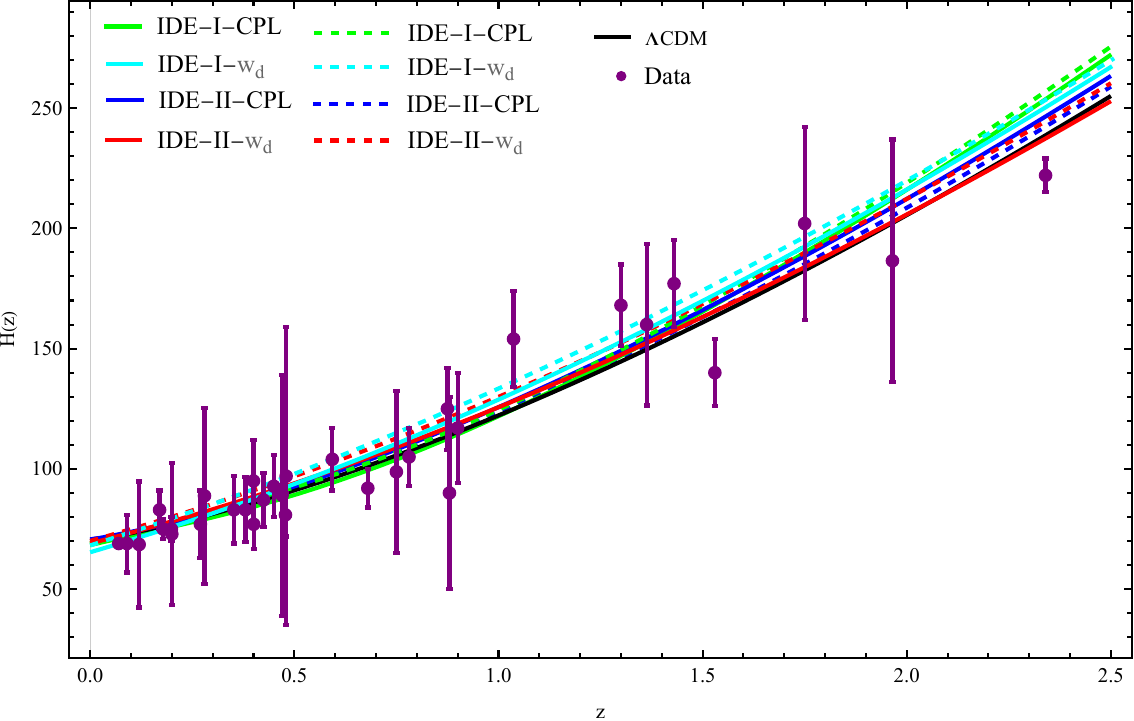}}
\caption{Left panel: comparison of the observational growth rate data points (see Table \ref{tab1}) and theoretical prediction of the growth rate $ f\sigma_{8}(z) $(see Eq.\ref{ffff2}) and explanations after it) as a function of redshift $ z$. Right panel: comparison of the 36 cosmic chronometer data points given in Table \ref{tabHdata} and the theoretical evolution of the Hubble parameter of the IDE models in terms of the redshift $ z$. The color lines and styles utilized in the inner panels of the figure are similar to those depicted in Fig.\ref{f7888}.}
\label{ffr7}
\end{figure*}
{\begin{equation}
 \Delta {T}(\%)=
\begin{cases}
 \text{{\small Homogenous }}\;\;\;\; \text{{\small Clustered}}\;\;\;\;\;\;\;\;\;\text{{\small Model}}\\
\sim  -2.31\%\; ; \;\;\;\; \sim -2.60\%\;\;\;\;\;\;\text{{\small IDE-I, CPL;}}\\
\sim  -1.52\%\; ;\;\;\;\; \sim -1.16\%\;\;\;\;\;\;\text{{\small IDE-I, }$w_{\mathrm{de}}$;}\\
\sim  -1.70\%\; ;\;\;\;\;\sim -2.12\%\;\;\;\;\;\;\text{{\small IDE-II, CPL;}}\\
\sim  -0.78\%\; ;\;\;\;\;\sim -0.48\%\;\;\;\;\;\;\text{{\small IDE-II, }$w_{\mathrm{de}}$}
\end{cases} \notag
\end{equation}}

In the subsequent step of our investigation, we focus on the growth of matter perturbations. 
This involves numerically solving Eqs.(\ref{ewq} $\& $ \ref{ecq}) for both homogeneous and clustered cases of DE in the context of IDE models.
To constrain the values of $\sigma_{8}$ and other free parameters associated with the IDE models, we perform a combined statistical analysis that incorporates background and growth rate data obtained from RSD (refer to Eq. (\ref{xi222}) and Subsec. \ref{growth}). The outcomes of this data analysis are presented in Table \ref{tabclus}.

With the obtained best-fit values listed in Table (\ref{tabclus}), we examine the evolution of the growth rate of matter perturbations, $f(z)$, and the percentage deviation $\Delta f(\%)$ compared to the $\Lambda\mathrm{CDM}$ model. 
The evolution of the linear growth rate of matter perturbations for different models as a function of redshift $z$ is displayed in the upper panel of Fig. \ref{f733}. The line $f=1$ corresponds to the Einstein-de Sitter (EdS) Universe, characterized by $\Omega_{\mathrm{dm}}=1$ and $\Omega_{\mathrm{de}}=0$. It is evident that as redshift increases, the linear growth rates of matter perturbations associated with all models approach and converge towards the constant EdS line.
We observe that the IDE-II model, incorporating both homogeneous and clustered DE, with the CPL parameterization and a constant value of $w_{\mathrm{de}}$, demonstrates a relatively smaller deviation from the evolution of the linear growth rate of matter perturbations observed in the $\Lambda$CDM model.

Furthermore, we can quantify the difference in the growth rate of the IDE models compared to its value in the $\Lambda \mathrm{CDM}$ model by calculating the percentage relative difference as follows:
\begin{equation}\label{deem}
\Delta f (\%)=100\times\Big[\frac{f_{\mathrm{model}}}{f_{\mathrm{\Lambda CDM}}}-1\Big]
\end{equation}
The $\Delta{{f}}(\%)$, as a function of redshift $z$ is illustrated in the lower panel of Fig. \ref{f733}. These values are calculated using the best-fit parameters provided in Table \ref{tabclus}. A positive (negative) $\Delta{{f}}$ indicates that the corresponding IDE models shows a higher (lower) linear growth rate of matter perturbations compared to the $\Lambda \mathrm{CDM}$ model. Listed below are the obtained results for $\Delta{{f}}(\%)$ at the present time for both homogeneous and clustered IDE models:
\begin{equation}
\Delta{f}(\%)=\begin{cases}
\text{{\small Homogenous }}\;\;\;\; \text{{\small Clustered }}\;\;\;\; \;\;\text{{\small Model}}\\
\sim  -1.58\% \; ; \;\;\; \sim -0.81\% \;\;\;\text{{\small IDE-I, CPL;}}\\
\sim  -1.38\% \; ; \;\;\; \sim -1.08\% \;\;\; \text{{\small IDE-I, }\;$w_{\mathrm{de}}$;}\\
\sim- 0.29\% \; ; \;\;\; \sim -0.10\% \;\;\; \text{{\small IDE-II, CPL;}}\\
\sim -0.19\% \; ; \;\;\; \sim -0.05\% \;\;\; \text{{\small IDE-II, $w_{\mathrm{de}}$;}}
\end{cases} \notag
\end{equation}
As illustrated in the lower panel of Fig. \ref{f733}, the evolution of the $\Delta{{f}}(\%)$ value is influenced by the clustering or homogeneity of DE, as well as the choice of the parameter for the EoS of DE.  In the IDE-I model, when the parameter considered is CPL, the $\Delta{{f}}(\%)$ value associated with the homogeneous DE surpasses the $\Delta{{f}}(\%)$ value related to the clustered DE at $ z\gtrsim 0.53 $. However, if a constant $w_{\mathrm{de}}$ is assumed for the EoS, the $\Delta{{f}}(\%)$value for the homogeneous DE is smaller compared to the clustered DE.
 Furthermore, in the IDE-II model, when we consider the CPL parameterization, $ \Delta{{f}}(\%)$ associated with clustered DE exceeds the value associated with homogeneous DE at $ z\lesssim1.40 $, and the opposite behavior is observed at $z\gtrsim 1.40  $. On the other hand, if we assume a constant value for $w_{\mathrm{de}}$, the $ \Delta{{f}}(\%)$ value for homogeneous DE is larger than that for clustered DE at $ z\gtrsim 0.3 $. 
 
 By examining the lower panel of Fig. \ref{f733} and the upper panel of Fig. \ref{f7888}, It is observed that when the $ \Delta \mathrm{E} $ of IDE models is positive, there is a corresponding negative $\Delta{{f}}$. This indicates an inverse relation between the evolution of $ \Delta \mathrm{E} $ and $\Delta{{f}}$. In other words, an increase in $ \Delta \mathrm{E} $ leads to a decrease in  $\Delta{{f}}$, and vice versa. Also, It is observed that when $ \Delta \mathrm{E} $  reaches its maximum value, $\Delta{{f}}$ is minimized.

Moreover, in the left panel of Fig. \ref{ffr7}, a comparison is presented between the observed data points (listed in Table \ref{tab1}) and the theoretical prediction of the growth rate of matter perturbations, $f\sigma_{8}(z)$,  (refer to Eq.(\ref{ffff2}) and the explanation after it). This analysis includes both homogeneous and clustered DE scenarios within the IDE models.
 \begin{table*}
  \centering
  \caption{The numerical results of model selection were conducted using background data for both Homogeneous (H) and Clustered (C) Dark Energy within the context of IDE models studied in this work.(see Eq. \ref{xi22}).}
  \begin{tabular}{lc|ccccccccc}
    \hline
    \hline
    Model &&$\mathrm{EoS}$ &$ k$ &$\chi^{2}_{\mathrm{min}}$&$\chi^{2}_{\mathrm{red}} $&$\mathrm{AIC}$&$ \mathrm{\Delta AIC}$&$ \mathrm{BIC} $&$\Delta  \mathrm{BIC} $\\ 
\hline
\hline
\multirow {2}{*}{IDE-I}
& {\footnotesize C} & {\footnotesize CPL} & 6 & 1075.51 &0.9848 &  1087.51 & 4.98 &   1117.51& 19.97 \\
& {\footnotesize H} & {\footnotesize CPL} & 6 & 1075.10 &  0.9845 &  1087.10 & 4.57 &  1117.10 & 19.56 \\
\hline
\multirow{2}{*}{IDE-I}
& {\footnotesize C} & $w_{\mathrm{d}}$& 5 & 1075.92 & 0.9843 & 1085.92&  3.39 & 1110.93& 13.39 \\
& {\footnotesize H} & $w_{\mathrm{d}}$ & 5 &1075.33 & 0.9838 & 1085.33 &2.82 & 1110.34& 12.80 \\
\hline
\multirow{2}{*}{IDE-II}
&{\footnotesize  C} &  {\footnotesize CPL} & 6 &  1075.82 & 0.9851&  1087.82 & 5.29 &  1117.83 &  20.29\\
& {\footnotesize H} &   {\footnotesize CPL} & 6 & 1075.41& 0.9847&  1087.41& 4.88 & 1117.42&19.88\\
\hline
\multirow{2}{*}{IDE-II}
 & {\footnotesize C} & $w_{\mathrm{d}}$ & 5 &  1075.91&  0.9844 &   1085.91 & 3.38 &  1110.92&13.38 \\
& {\footnotesize H} &  $w_{\mathrm{d}}$ & 5 & 1075.45& 0.9839 &  1085.45 & 2.92 & 1110.46&12.92\\
\hline
$\Lambda$CDM & - & - & 3 & 1076.53 &  0.9840&1082.53& --- & 1097.54&--- \\
\hline
\hline
\end{tabular}
\label{tabbacgrowth}
\end{table*}
\begin{table*}
  \centering
  \caption{ The numerical outcomes of model selection based on background and growth rate data jointly for both Homogeneous (H) and Clustered (C) Dark Energy in the context of IDE models studied in this work(see Eq. \ref{xi222}).}
\begin{tabular}{lc|cccccccc}
\hline
\hline
 Model &&$\mathrm{EoS}$ &\;$ k$ &$\chi^{2}_{\mathrm{min}}$\;&$\chi^{2}_{\mathrm{red}} $\;&$\mathrm{AIC}$&$ \mathrm{\Delta AIC}$&$ \mathrm{BIC} $&$\Delta  \mathrm{BIC} $\\ 
\hline
\hline
\multirow{2}{*}{IDE-I}
& {\footnotesize C} & {\footnotesize CPL} & 7 & 1108.25 & 0.9764& 1122.25 & 3.93 & 1157.48&19.02 \\
& {\footnotesize H} & {\footnotesize CPL} & 7 & 1108.81 & 0.9769 &1122.81& 4.49 & 1158.05& 19.57 \\
\hline
\multirow{2}{*}{IDE-I}
&{\footnotesize  C} & $w_{\mathrm{d}}$ & 6 &1109.24 & 0.9764 &1121.24 & 2.92 & 1151.44& 12.96 \\
& {\footnotesize H} &  $w_{\mathrm{d}}$ &6 &1109.43 & 0.9766& 1121.43 & 3.11 & 1151.63& 13.15 \\
\hline
\multirow{2}{*}{IDE-II}
& {\footnotesize C} & {\footnotesize CPL }& 7 & 1108.51 & 0.9765 &1122.51 & 4.19 &1157.75&19.27\\
& {\footnotesize H} & {\footnotesize CPL} & 7 & 1108.72 &0.9768 & 1122.72 &4.40  & 1157.95 & 19.47\\
\hline
\multirow{2}{*}{IDE-II}
 & {\footnotesize C} & $w_{\mathrm{d}}$ &6 & 1109.61 &0.9767& 1121.61& 3.29 & 1151.81&13.33 \\
& {\footnotesize H} &  $w_{\mathrm{d}}$ & 6 & 1109.47 & 0.9766 &  1121.47 & 3.15 & 1151.67&13.19 \\
\hline
$\Lambda$CDM & -- & -- & 4 &1110.32 & 0.9756 &1118.32 & -- &1138.48&-- \\
\hline
\hline
\end{tabular}
\label{tabcluss}
\end{table*}
\begin{table*}
 \centering
 \caption{The numerical outcomes of parameter fitting with a $1\sigma $ confidence level, are obtained for the IDE models examined in this study for both Homogeneous(H) and Clustered(C) DE. These results are obtained from the combination of a dataset, as defined in Eq. (\ref{xi22}), which includes background dataset, BAO+ SnIa+CMB +H.}
 \begin{tabular}{lc|c|cccccccc}
\hline
\hline
Model &&$\mathrm{EoS}$ &$\Omega_{b}h^{2}$ & $\Omega_{c}h^{2}$&$ \Omega_{m0} $&$H_{0}$&$ w_{0}$&$ w_{1} $&$ \xi_{1} $\\
\hline
\multirow{2}{*}{IDE-I}
& {\footnotesize C}&{\footnotesize CPL}&{$ 0.02190^{+0.045}_{-0.052}$}&
{$0.1103^{+0.023}_{-0.034}$}&{$0.2833^{+ 0.88}_{- 0.93}$}&
{$68.34^{+0.721}_{-0.789}$}&{$-0.640^{+0.096}_{-0.11}$}&
{$-1.61^{+0.604}_{-0.405}$}&{$ -0.2458^{+0.017}_{-0.034}$}\\
&{\scriptsize H}&{\footnotesize CPL}&{$  0.02198^{+0.017}_{-0.015}$}&
{$0.1074^{+0.016}_{-0.007}$}&{$0.2799^{+0.014}_{-0.009}$}&
{$67.96^{+1.132}_{-1.713}$ }&{$ -0.810^{+ 0.088}_{- 0.065}$}&
{$-1.97^{+ 0.042}_{- 0.085}$}&{$ -0.2468^{+0.040}_{-0.032}$}\\
\hline
\hline
&&&$\Omega_{b}h^{2}$ & $\Omega_{c}h^{2}$&$ \Omega_{m0} $&$H_{0}$&$ w_{\mathrm{d}}$&$  \xi_{1}$ \\
\hline
\multirow{2}{*}{IDE-I}
&{\scriptsize C}& $ w_{\mathrm{d}}$&{$0.0222^{+0.017}_{-0.032}$}&
{$0.1072^{+0.031}_{-0.015}$}&{$ 0.2760^{+ 0.063}_{- 0.084}$}&
{$68.51^{+0.851}_{-0.919}$}&{$-1.102^{+0.063}_{-0.062}$}&
{$-0.1980^{+0.014}_{-0.016}$}&$ -- $\\
& {\scriptsize H}&$ w_{\mathrm{d}}$&{$ 0.02091^{+0.002}_{-0.011}$}&
{$0.1088^{+0.017}_{-0.014}$}&{$0.2790^{+0.023}_{-0.035}$}&
{$68.21^{+1.310}_{-1.521}$ }&{$-1.050^{+ 0.048}_{- 0.043}$}&
{$-0.1960^{+ 0.110}_{- 0.205}$}&$ -- $\\
\hline
\hline
&&&$\Omega_{b}h^{2}$ & $\Omega_{c}h^{2}$&$ \Omega_{m0} $&$H_{0}$&$ w_{0}$&$ w_{1} $&$ \xi_{2} $& \\
 \hline
\multirow{2}{*}{IDE-II}  
& {\scriptsize C} &{\footnotesize CPL}&$0.02301^{+0.027}_{-0.031}$&
 $0.1231^{+0.041}_{-0.052}$ &{$ 0.2950^{+0.051}_{-0.044}$}&
{$70.63^{+ 0.056}_{- 0.082}$}&{$-0.830^{+0.041}_{-0.059}$}&
{$0.190^{+0.063}_{-0.075}$}&{$-0.0260^{+0.008}_{-0.006}$}\\
 &{\scriptsize H} &{\footnotesize CPL}&{$ 0.02304^{+0.003}_{-0.008}$}&
{$0.1223^{+0.004}_{-0.009}$}&{$0.2942^{+0.013}_{-0.017}$}&
{$70.71^{+1.515}_{-1.613}$ }&{$-0.890^{+0.096}_{-0.121}$}&
{$  0.141^{+ 0.012}_{- 0.025}$}&{$ -0.0274^{+0.013}_{-0.021}$}\\
\hline
\hline
&&&$\Omega_{b}h^{2}$ & $\Omega_{c}h^{2}$&$ \Omega_{m0} $&$H_{0}$&$ w_{\mathrm{d}}$&$ \xi_{2}$ \\
\hline
\multirow{2}{*}{IDE-II}  
&{\scriptsize C} &$ w_{\mathrm{d}}$&{$0.01838^{+0.027}_{-0.033}$}&
{$0.1121^{+0.031}_{-0.064}$}&{$0.2740^{+ 0.081}_{- 0.077}$}&
{$69.04^{+0.751}_{-0.819}$}&{$-0.878^{+0.071}_{-0.095}$}&
{$0.0163^{+0.005}_{-0.009}$}&$ -- $\\
&{\scriptsize H}&$ w_{\mathrm{d}}$&{$ 0.01985^{+0.008}_{-0.009}$}&
{$0.1132^{+0.011}_{-0.013}$}&{$0.2716^{+0.073}_{-0.095}$}&
{$70.03^{+1.642}_{-1.743}$ }&{$ -0.95^{+ 0.178}_{- 0.241}$}&
{$0.0180^{+ 0.003}_{- 0.025}$}&$ -- $\\
\hline
\multirow{1}{*}{$\Lambda\mathrm{CDM}$}&&$ -1 $
&{$0.0244^{+ 0.011}_{- 0.014} $  }&{$0.1249 ^{+0.006}_{-0.009} $}&
{$0.2993^{+0.057}_{-0.063} $  }&{$69.96^{+0.273}_{-0.284} $}&
{$-1  $}&{$--$ }&{$--$ }&\\
\hline
\end{tabular}\label{tabhom}
\end{table*}
\begin{table*}
 \centering
 \caption{Numerical results of parameter fitting with 1$\sigma $ confidence level in the IDE models investigated in this work assuming Homogeneous (H) and Clustered (C) DE from combining the data set using Eq.(\ref{xi222}) based on background and growth rate data, BAO+SnIa+CMB +H+RSD.}
 \begin{tabular}{lc|c|ccccccccc}
\hline
\hline
Model &&$\mathrm{EoS}$ &$\Omega_{b}h^{2}$ & $\Omega_{c}h^{2}$&$ \Omega_{m0} $&$H_{0}$&$ w_{0}$&$ w_{1} $&$ \xi_{1} $&$\sigma_{8}$& \\
\hline
\multirow{2}{*}{IDE-I}  
&{\scriptsize C}&{\footnotesize CPL}&{$0.02131^{+0.011}_{-0.013}$}&
{$0.1212^{+0.020}_{-0.023}$}&{$0.2931^{+ 0.063}_{- 0.073}$}&
{$69.75^{+0.731}_{-0.719}$}&{$-0.89^{+0.063}_{-0.095}$}&
{$-1.38^{+0.114}_{-0.117}$}&{$ -0.239^{+0.057}_{-0.058}$}&
{$0.830^{+0.087}_{-0.084}$}\\
&{\scriptsize H}&{\footnotesize CPL}&{$  0.02194^{+0.016}_{-0.024}$}&
{$0.1157^{+0.025}_{-0.033}$}&{$0.2910^{+0.093}_{-0.085}$}&
{$68.80^{+1.212}_{-1.324}$ }&{$ -0.71^{+ 0.092}_{- 0.065}$}&
{$-1.76^{+ 0.322}_{- 0.415}$}&{$ -0.249^{+0.043}_{-0.061}$}&
{$ 0.814^{+0.093}_{-0.086}$}\\
\hline
\hline
&&&$\Omega_{b}h^{2}$ & $\Omega_{c}h^{2}$&$ \Omega_{m0} $&$H_{0}$&$ w_{\mathrm{d}}$&$  \xi_{1}$&$\sigma_{8}$ \\
\hline
\multirow{2}{*}{IDE-I}  
&{\scriptsize C}&$ w_{\mathrm{d}}$&{$0.02032^{+0.017}_{-0.013}$}&
{$0.1050^{+0.011}_{-0.034}$}&{$0.2740^{+ 0.072}_{- 0.083}$}&
{$67.66^{+0.741}_{-0.719}$}&{$-0.69^{+0.053}_{-0.074}$}&
{$-0.207^{+0.114}_{-0.117}$}&{$ 0.809^{+0.137}_{-0.154}$}&$ -- $\\
&{\scriptsize H}&$ w_{\mathrm{d}}$&{$0.01135^{+0.012}_{-0.014}$}&
{$0.1102^{+0.015}_{-0.013}$}&{$0.2691^{+0.043}_{-0.085}$}&
{$67.21^{+1.112}_{-1.125}$ }&{$ -0.61^{+ 0.028}_{- 0.065}$}&
{$-0.209^{+ 0.212}_{- 0.415}$}&{$ 0.799^{+0.133}_{-0.161}$}&$ -- $\\
\hline
\hline
&&&$\Omega_{b}h^{2}$ & $\Omega_{c}h^{2}$&$ \Omega_{m0} $&$H_{0}$&$ w_{0}$&$ w_{1} $&$ \xi_{2} $&$\sigma_{8}$& \\
 \hline
\multirow{2}{*}{IDE-II}  
&{\scriptsize C}&{\footnotesize CPL}&{$0.02126^{+0.017}_{-0.013}$}
 &{$0.1204^{+0.027}_{-0.040}$}&{$ 0.2877^{+0.031}_{-0.019}$}&
{$70.21^{+ 0.033}_{- 0.073}$}&{$-1.03^{+0.021}_{-0.019}$}&
{$0.350^{+0.043}_{-0.075}$}&{$-0.028^{+0.034}_{-0.039}$}&
{$ 0.810^{+0.062}_{-0.094}$}\\
&{\scriptsize H} &{\footnotesize CPL}&{$0.02178^{+0.016}_{-0.019}$}&
{$0.1153^{+0.025}_{-0.033}$}&{$0.2813^{+0.083}_{-0.095}$}&
{$69.83^{+1.311}_{-1.410}$ }&{$-1.08^{+ 0.068}_{- 0.095}$}&
{$ 0.381^{+ 0.212}_{- 0.325}$}&{$ -0.023^{+0.043}_{-0.052}$}&
{$0.820^{+0.063}_{-0.082}$}\\
\hline
\hline
&&&$\Omega_{b}h^{2}$ & $\Omega_{c}h^{2}$&$ \Omega_{m0} $&$H_{0}$&$ w_{\mathrm{d}}$&$  \xi_{2}$&$\sigma_{8}$ \\
\hline
\multirow{2}{*}{IDE-II}  
&{\scriptsize C}&$ w_{\mathrm{d}}$&{$0.02169^{+0.036}_{-0.041}$}&
{$0.1112^{+0.041}_{-0.021}$}&{$0.2821^{+ 0.043}_{- 0.073}$}&
{$68.61^{+0.624}_{-0.879}$}&{$-0.81^{+0.063}_{-0.075}$}&
{$0.015^{+0.013}_{-0.017}$}&{$0.819^{+0.088}_{-0.091}$}&$ -- $\\
&{\scriptsize H} & $ w_{\mathrm{d}}$&{$0.02068^{+0.016}_{-0.019}$}&
{$0.1121^{+0.015}_{-0.013}$}&{$0.2835^{+0.063}_{-0.085}$}&
{$68.47^{+1.012}_{-1.053}$ }&{$-0.76^{+ 0.068}_{- 0.085}$}&
{$0.013^{+ 0.012}_{- 0.015}$}&{$ 0.811^{+0.103}_{-0.109}$}&$ -- $\\
 \hline
\multirow{1}{*}{$\Lambda\mathrm{CDM}$}&&-1
&{$0.0244^{+ 0.011}_{- 0.014} $  }   & 
{$0.1249 ^{+0.006}_{-0.009} $   }  &
{$0.2993^{+0.057}_{-0.063} $  }  &
{$69.96^{+0.273}_{-0.284} $  } &
{$-1  $   }                                    &
{$--  $  }                                     &
{$0.752^{+ 0.022}_{- 0.022} $ }&$  --$\\ 
\hline
\end{tabular}\label{tabclus}
\end{table*}
\begin{figure*}
\centering
\setlength\fboxsep{0pt}
\setlength\fboxrule{0pt}
\fbox
{\includegraphics[width=8.6cm]{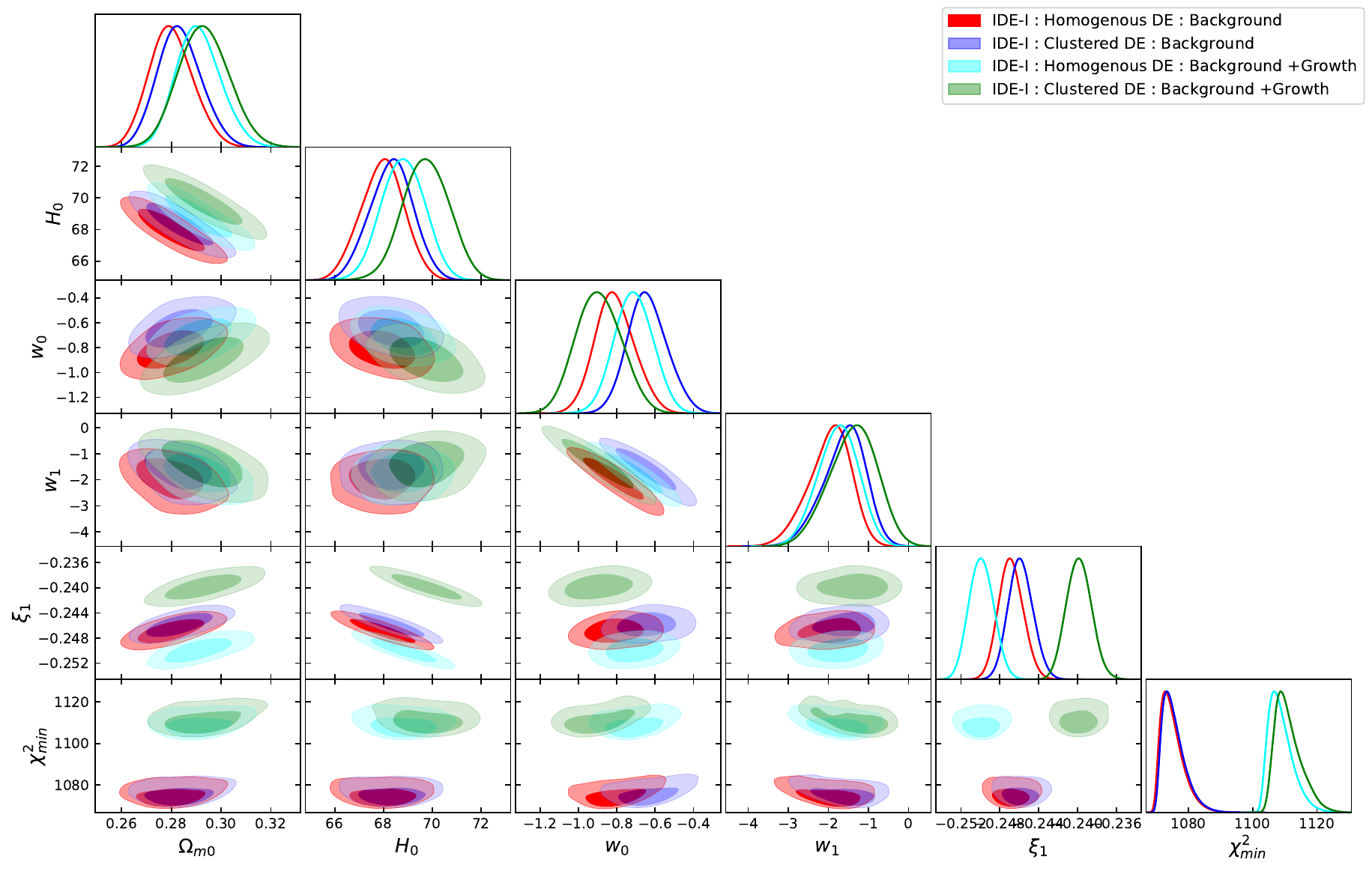}}
{\includegraphics[width=8.6cm]{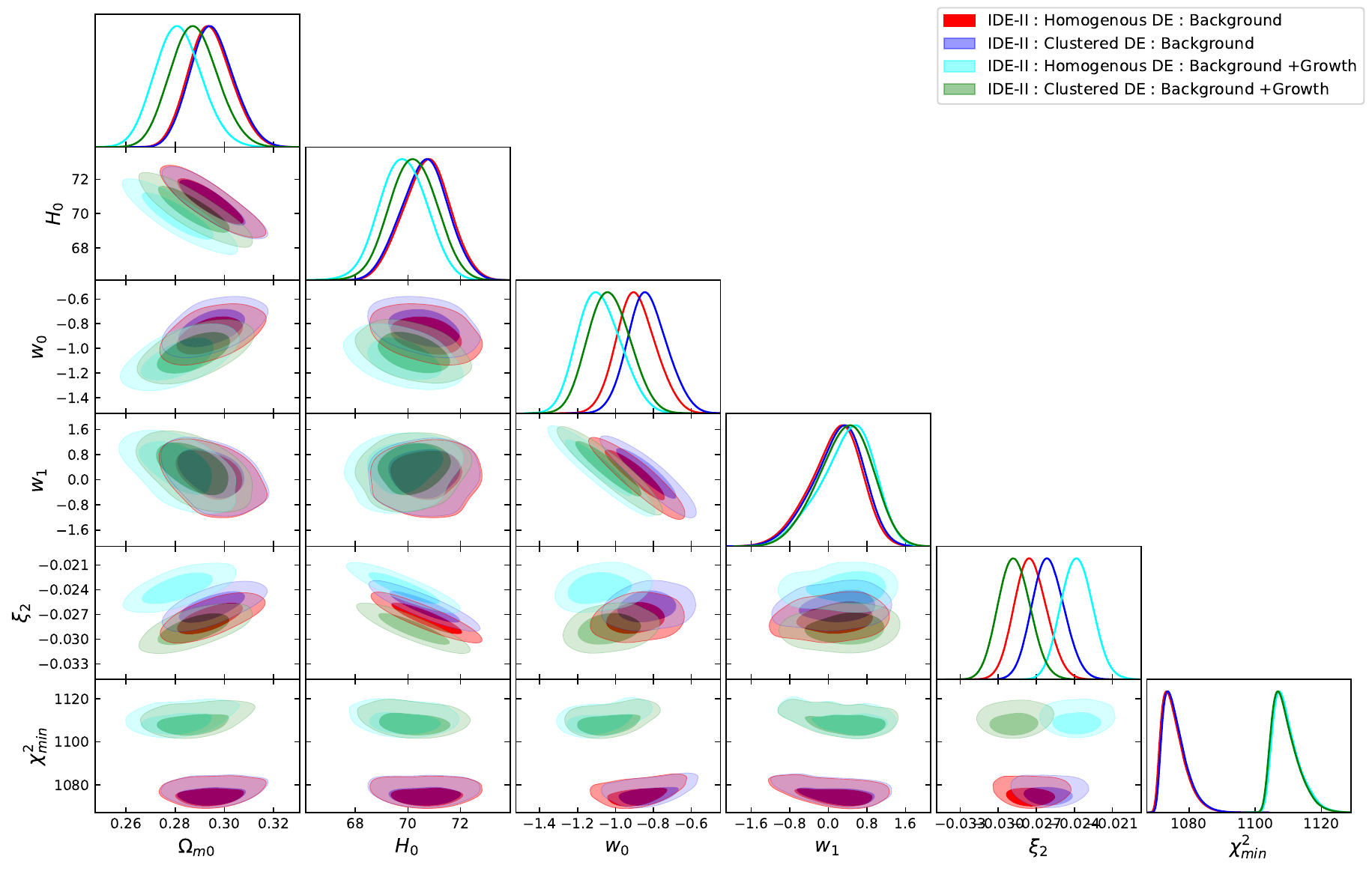}}
{\includegraphics[width=8.6cm]{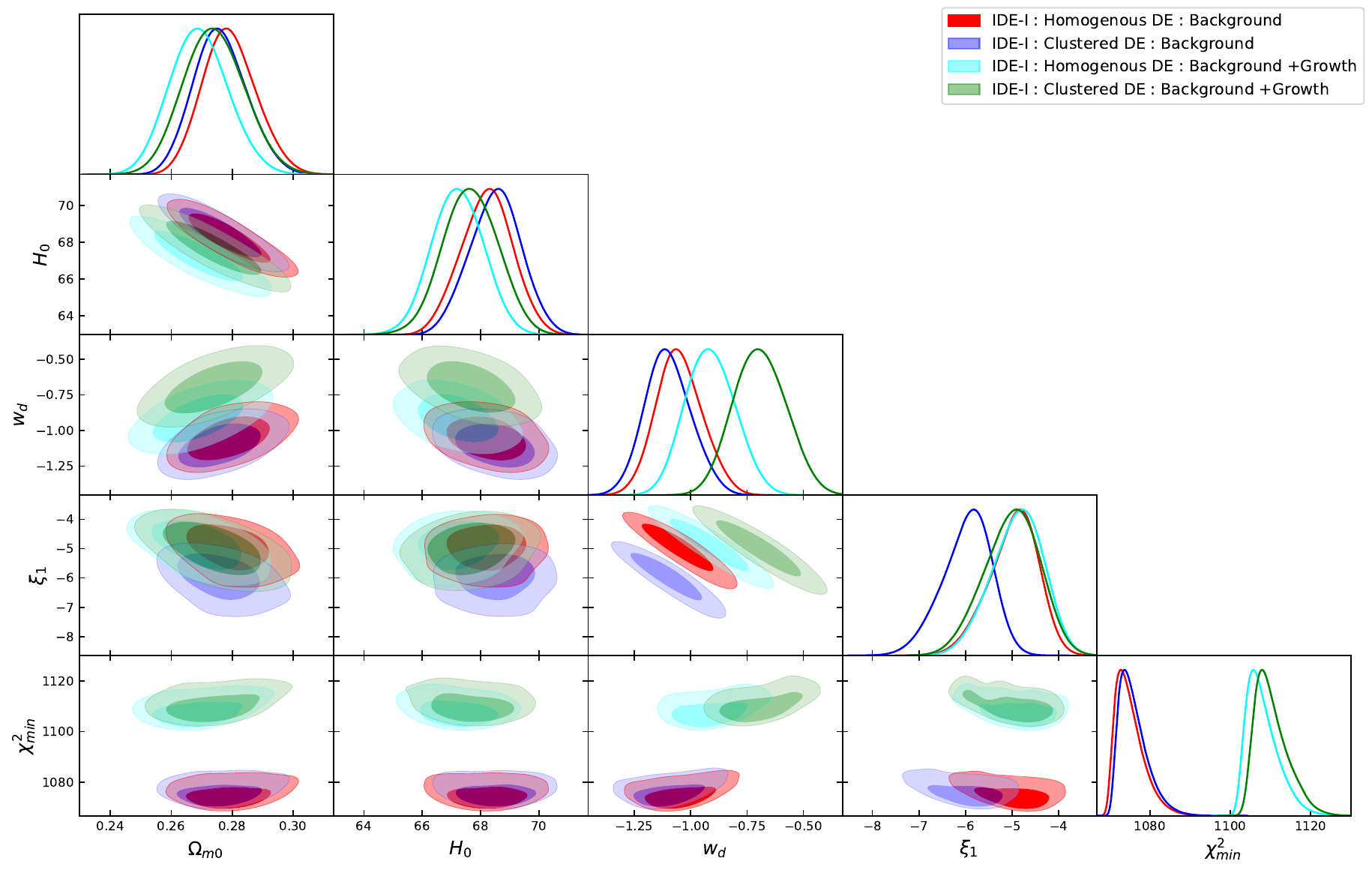}}
{\includegraphics[width=8.6cm]{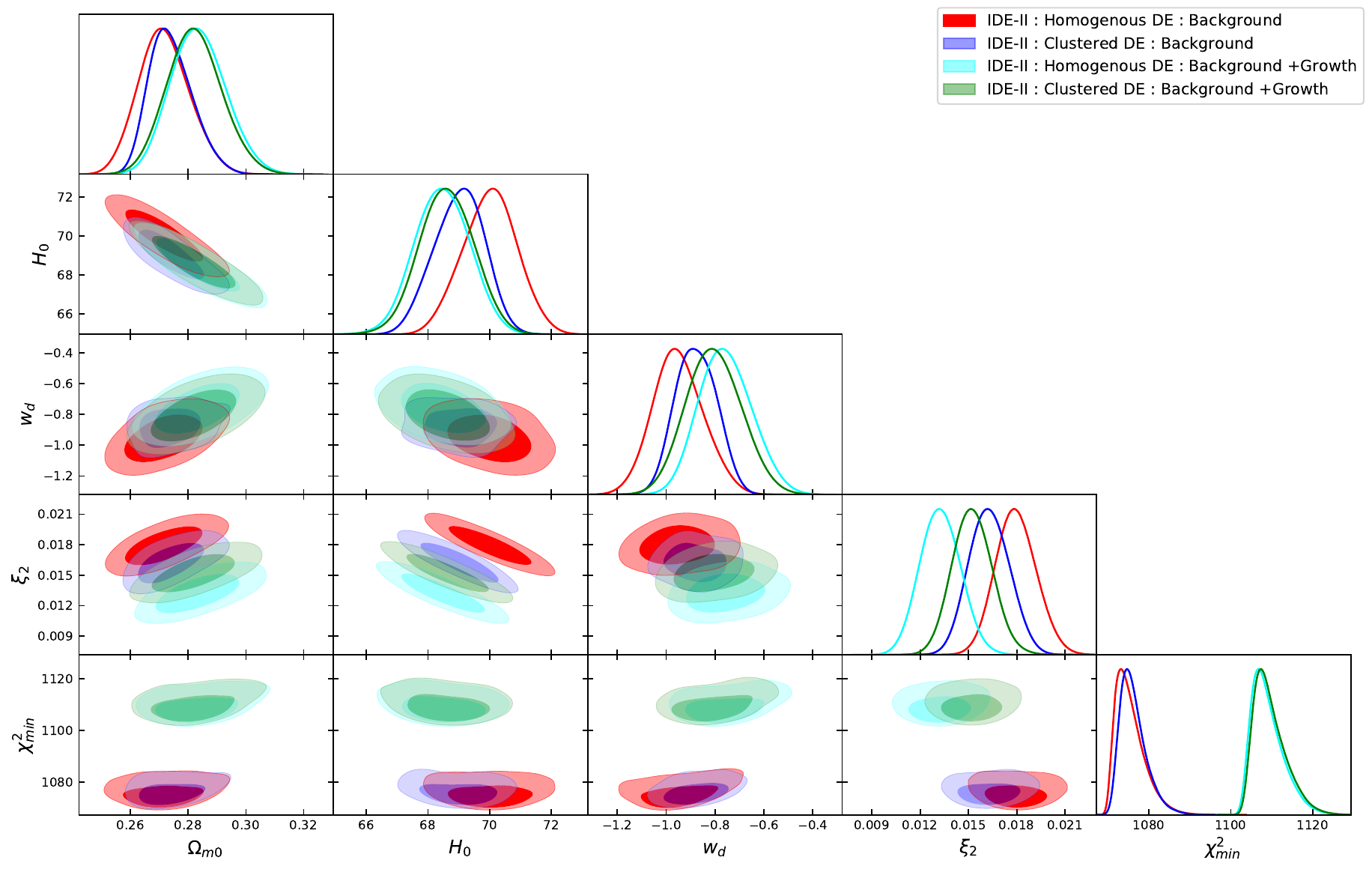}}

\caption{The confidence levels of the $1\sigma$ and $2\sigma$ limits for the IDE-I and IDE-II models. The upper panels depict the EoS parameter of DE using the CPL parameterization for both IDE-I (upper left panel) and IDE-II (upper right panel) models. The lower panels focus on the fixed EoS parameter, $w_{\mathrm{d}}$, for both IDE-I (lower left panel) and IDE-II (lower right panel) models.
These confidence levels are determined using the background dataset alone for both homogeneous (red) and clustered DE (blue) scenarios. Additionally, the combined background and growth rate dataset is utilized for both homogeneous (cyan) and clustered DE (green) scenarios.
For more details, refer to Eqs.(\ref{xi22} \& \ref{xi222}), as well as Table \ref{tabclus} for numerical values.}
\label{f777}
\end{figure*}
\subsection{\textbf{ AIC and BIC Criteria}}
When comparing models, the $\chi^{2}_{\mathrm{min}}$ can be used if the models have the same degrees of freedom. In this case, a smaller $\chi^{2}_{\mathrm{min}}$ indicates a better fit to the observational data. 
If the degrees of freedom are not equal, the reduced chi-square statistic $ \chi^{2}_{\mathrm{red}} = \chi^{2}_{\mathrm{min}} / (N - k) $ can be used, where $ k $ is the number of free parameters in the model and $ N $ is the total number of data points. When $ \chi^{2}_{\mathrm{red}} $ is around 1, it suggests a good fit to the data. Values significantly smaller or larger than 1 ($ \chi^{2}_{\mathrm{red}}\ll 1 $ or $ \chi^{2}_{\mathrm{red}}\gg 1 $) indicate that the model is not desirable and should be discarded.

Additionally, the Akaike Information Criterion(AIC) \citep{ Akaike1974} and Bayesian Information Criterion(BIC)\citep{Schwarz1978} can be used to select the most appropriate model based on its compatibility with the observational data. The AIC and BIC are defined as: $\mathrm{AIC} = -2\ln \mathcal{L}_{\mathrm{max}}+ 2k$ and $\mathrm{BIC} = -2\ln \mathcal{L}_{\mathrm{max}} + k\ln N$, where $ \mathcal{L}_{\mathrm{max}} $ represents the maximum value of the likelihood, which is related to $ \chi^{2}_{\mathrm{min}} $ as $ \chi^{2}_{\mathrm{min}} = -2\ln \mathcal{L}_{\mathrm{max}} $. Both the AIC and BIC consider the number of free parameters (k) and the total number of data points (N). In this case, N specifically refers to 1098 data points for the background data and expands to 1142 when accounting for both the background and growth data jointly (see Sec. \ref{sec3}). By calculating the differences between the AIC and BIC of models and a reference model (often chosen as the best-fitting model), we can assess the relative support for each model.
The differences $\Delta\mathrm{AIC}$ and $\Delta \mathrm{BIC}$ are calculated as follows\citep{Rivera2019, Liddle2007}:
\begin{align}\label{Aic}
\Delta \mathrm{AIC}&=\mathrm{AIC}_{\mathrm{model}}-\mathrm{AIC}_{\Lambda \mathrm{CDM}}=  \Delta \chi^{2}_{\mathrm{min}}+2\Delta k\\ \notag
\Delta \mathrm{BIC}&=\mathrm{BIC}_{\mathrm{model}}-\mathrm{BIC}_{\Lambda \mathrm{CDM}}=  \Delta \chi^{2}_{\mathrm{min}}+\Delta k(\ln \mathrm{N})
\end{align}
The interpretations of these criteria suggest different levels of support or evidence against a model based on the magnitudes of $ \Delta \mathrm{AIC} $ and $ \Delta \mathrm{BIC} $. For example, small values of $ \Delta \mathrm{AIC} $ or $ \Delta \mathrm{BIC} $ indicate substantial support or weak evidence against the model, respectively. Larger values of $ \Delta \mathrm{BIC} $ indicate better agreement with the observational data.
In summary, the $\chi^{2}_{\mathrm{min}}$, AIC, and BIC are used to compare models. The choice of which criterion to use depends on the degrees of freedom and the emphasis placed on goodness of fit versus model complexity.
In \citep{Rivera2019}, guidelines are presented to assess model support using $ \Delta \mathrm{AIC}$ and $ \Delta \mathrm{BIC}$. Substantial support is given when $|\Delta \mathrm{AIC}|$ is in (0, 2], while considerably less support is in [4, 7]. Models with $|\Delta \mathrm{AIC}|$ exceeding 10 are considered inappropriate. Similar criteria apply to $|\Delta \mathrm{BIC}|$, indicating weak, positive, strong, or very strong evidence against the model. Larger $\Delta \mathrm{BIC}$ values suggest better consistency with the observational data.

We present the computed results in Tables (\ref{tabbacgrowth} \& \ref{tabcluss}). These tables are obtained using the numerical values from Tables (\ref{tabhom} \& \ref{tabclus}), taking into account the CPL parametrization and a constant $w_{\mathrm{de}}$ for the investigated IDE models. Also, the analysis assumes both the homogeneity and clustering of DE.
 
According to the analysis of AIC and BIC, it can be inferred that the selection of a model that is more consistent with the observational data (including background and growth rate data) depends on two factors: the homogeneity or clustering of DE and the EoS parameter of DE ( Specifically, in this study, the CPL parametrization and a constant $w_{\mathrm{de}}$). Moreover, the model selection also depends on the dataset utilized.

For instance, if we perform AIC analysis and focus solely on the background data, we can deduce that IDE-I($w_{\mathrm{de}}$) and IDE-II($w_{\mathrm{de}}$) models demonstrate a higher level of compatibility with the observational data compared to the other models. This holds for both homogeneous and clustered DE. However, when considering homogeneous DE, there is a slightly better agreement with the observational data(see Table \ref{tabbacgrowth}).

Moreover, when conducting AIC analysis and considering both the background and growth rate data simultaneously, it can be inferred that models IDE-I($w_{\mathrm{de}}$) and IDE-II($w_{\mathrm{de}}$) show greater compatibility with the observational data compared to the other models. This finding holds true for both homogeneous and clustered DE. Notably, when specifically examining clustered DE, model IDE-I($w_{\mathrm{de}}$) demonstrates a slightly better agreement with the observational data in comparison to homogeneous DE (see Table \ref{tabcluss}).

In summary, when utilizing the background data, the AIC analysis shows that for homogeneous DE, the models IDE-I(wd), IDE-II(wd), IDE-I(CPL), and IDE-II(CPL) exhibit a better fit with the observational data, respectively. Conversely, when considering clustered DE, the models IDE-II(wd), IDE-I(wd), IDE-I(CPL), and IDE-II(CPL) demonstrate better compatibility with the data (see Table \ref{tabbacgrowth}). 
Also, when using the background and growth data jointly, the assumption of either homogeneous or clustered DE not only affects the fitting of the models to the observational data but also modifies the order in which the models fit the observational data.

\section{Conclusions}\label{conclude}
In this study, we utilized a two-step approach to investigate two well-known IDE models, considering two distinct cases for the EoS parameter of DE (CPL parameterization and a constant value for $ w_{\mathrm{de}} $).

Firstly, we performed an MCMC analysis to constrain the free parameters of the models based on the latest available background data (see Sec. \ref{sec3} and Eq. (\ref{xi22})). The results of our data analysis pertaining to the IDE models were summarized in Table \ref{tabhom}. Additionally, the left panels of Fig. \ref{f732q} illustrated the confidence levels for $1\sigma$ and $2\sigma$ constraints on the IDE models based on the background datasets. These triangular plots visually depicted the correlations between each pair of free parameters in the models.

Following that, we utilized the best-fit values obtained from the data analysis to investigate significant background parameters such as $\Delta E$, $q$, and $\Delta T$. These parameters were examined to compare the models with each other as well as with the $\Lambda\mathrm{CDM}$ model.

Concerning the Hubble parameter, we concluded that the IDE-I model, with the CPL parameterization and a constant value for $w_{\mathrm{de}}$, exhibited negative values of $\Delta E(\%)$ compared to the $\Lambda \mathrm{CDM}$ model for all redshifts. This result is true for both homogeneous and clustered DE scenarios (see top panel of Fig. \ref{f7888}).
 For the IDE-II model with the same parameterization, positive values of $\Delta E(\%)$ were obtained at specific redshift ranges. These findings indicate that the cosmic expansion in the IDE models can be either larger or smaller than the $\Lambda \mathrm{CDM}$ model, depending on the specific model and parameters. Additionally, a comparison between the theoretical evolution of the Hubble parameter and cosmic chronometer data was presented in the right panel of Fig. \ref{ffr7}.

Moreover, the calculated transition time from the decelerated expansion phase ($q > 0$) to the accelerated expansion phase ($q < 0$) in the studied IDE models was found to be comparable to the transition time obtained in the $\Lambda\mathrm{CDM}$ model. This comparison is presented in the middle panel of Fig. \ref{f7888}.

In addition to the previously mentioned quantities, we also calculated the age of the Universe within each of the IDE models. Interestingly, we observed that the age of the Universe in the IDE-II ($w_{\mathrm{de}}$) models exhibited better comparability to the age of the Universe in the standard $\Lambda\mathrm{CDM}$ model, for both the homogeneous and clustered DE scenarios. The evolution of $\Delta T(\%)$ as a function of redshift ($z$) is presented in the lower panel of Fig. \ref{f7888}.

In the subsequent step of our investigation, we focused on matter perturbation growth in the context of IDE models. We solved the relevant equations numerically for both homogeneous and clustered cases of DE. To constrain the parameters of the IDE models, including $\sigma_8$, we performed a combined statistical analysis using background and growth rate data obtained from RSD. The results are summarized in Table \ref{tabclus}. 

Using the best-fit values from Table \ref{tabclus}, we analyzed the evolution of the growth rate of matter perturbations, $ f(z)$ and its deviation, $\Delta f(\%)$ from the $\Lambda\mathrm{CDM}$ model. 
The lower panel of Fig. \ref{f733} displays $\Delta f(\%)$ as a function of redshift. Positive (negative) values indicate higher (lower) growth rates compared to the $\Lambda\mathrm{CDM}$ model. The clustering or homogeneity of DE, as well as the choice of the parameter for the EoS of DE, influence the evolution of $\Delta f(\%)$. 

In the IDE-I model with the CPL parameterization, we concluded that the $\Delta f(\%)$ value for homogeneous DE surpasses the value for clustered DE at $z\gtrsim0.53$. However, assuming a constant $w_{\mathrm{de}}$ for the EoS, the $\Delta f(\%)$ value for homogeneous DE is smaller than that for clustered DE. 
Also, In the IDE-II model with the CPL parameterization, the $\Delta f(\%)$ value related to the clustered DE exceeds the value for homogeneous DE at $z\lesssim1.40$, while the opposite behavior is observed at $z\gtrsim1.40$. Assuming a constant $w_{\mathrm{de}}$, the $\Delta f(\%)$ value for homogeneous DE is larger than that for clustered DE at $z\gtrsim0.3$. 

Following that, a comparison was conducted between the growth rate data (Table \ref{tab1}) and the theoretical prediction of the growth rate, $f \sigma_{8}(z)$. we observed that the IDE models demonstrated good compatibility with the growth rate data (refer to the left panel of Fig. \ref{ffr7}). Additionally, in panels of Figs. \ref{f777} and \ref{f732q} , we illustrated the confidence levels representing the 1$\sigma$ and 2$\sigma$ constraints on the IDE models for both homogeneous and clustered DE. These constraints were determined through an analysis of background and growth rate data.

Eventually, the analysis of AIC and BIC revealed that the selection of a model consistent with the observational data depended on the homogeneity or clustering of DE and the EoS parameter of DE (specifically, CPL parametrization and a constant $w_{\mathrm{de}}$). The choice of dataset also influenced the model selection.

For background data analysis alone, IDE-I($w_{\mathrm{de}}$) and IDE-II($w_{\mathrm{de}}$) models demonstrated higher compatibility with the observational data, regardless
 of homogeneity or clustering of DE. However, homogeneous DE showed slightly better agreement with the data.

Considering both background and growth rate data, IDE-I($w_{\mathrm{de}}$) and IDE-II($w_{\mathrm{de}}$) models exhibited greater compatibility with the observational data for both homogeneous and clustered DE. However, IDE-I($w_{\mathrm{de}}$) showed slightly better agreement with the data in the case of clustered DE.

In summary, when utilizing the background data, the AIC analysis indicated that for homogeneous DE, the models IDE-I(wd), IDE-II(wd), IDE-I(CPL), and IDE-II(CPL) provided a better fit with the observational data, respectively. In adition, for the clustered DE, the models IDE-II(wd), IDE-I(wd), IDE-I(CPL), and IDE-II(CPL) demonstrated better compatibility with the data (see Table \ref{tabbacgrowth}). Also, when using the background and growth data jointly,
the assumptions of homogeneity or clustering of DE not only affected the model fitting to the data but also modified the order in which the models fit the observational data (see Table \ref{tabcluss}).
\twocolumngrid
\bibliographystyle{IEEEtran}
\bibliography{wimp}
\end{document}